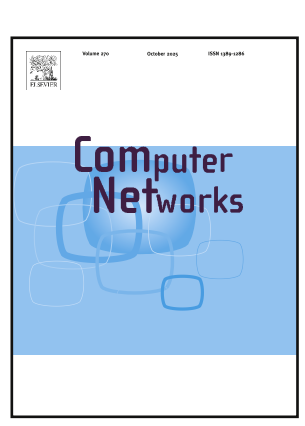

# Sharpening Kubernetes audit logs with context awareness

Matteo Franzil [a,b,*], Valentino Armani [b], Luis Augusto Dias Knob [b], Domenico Siracusa [a]

[a] Department of Information Engineering and Computer Science, University of Trento, Via Sommarive 9, Trento, 38123, Italy
[b] Center for Cybersecurity, Fondazione Bruno Kessler, Via Sommarive 18, Trento, 38123, Italy

ABSTRACT

Kubernetes (K8s) has emerged as the de facto orchestrator of microservices, providing scalability and extensibility to a highly dynamic environment. It builds an intricate and deeply connected system that requires extensive monitoring capabilities to be properly managed. To this account, K8s natively offers audit logs, a powerful feature for tracking Application Programming Interface (API) interactions in the cluster. Audit logs provide a detailed and chronological record of all activities in the system. Unfortunately, K8s auditing suffers from several practical limitations: it generates large volumes of data continuously, as all components within the cluster interact and respond to user actions. Moreover, each action can trigger a cascade of secondary events dispersed across the log, with little to no explicit linkage, making it difficult to reconstruct the context behind user-initiated operations.

In this paper, we introduce K8NTEXT, a novel approach for streamlining K8s audit logs by reconstructing *contexts*, i.e., grouping actions performed by actors on the cluster with the subsequent events these actions cause. Correlated API calls are automatically identified, labeled, and consistently grouped using a combination of inference rules and a Machine Learning (ML) model, largely simplifying data consumption. We evaluate K8NTEXT's performance, scalability, and expressiveness both in systematic tests and with a series of use cases. We show that it consistently provides accurate context reconstruction, even for complex operations involving 50, 100 or more correlated actions, achieving over 95 % accuracy across the entire spectrum, from simple to highly composite actions.

## 1. Introduction

Kubernetes (K8s) is a powerful and flexible platform for managing containerized applications. It is designed to be highly scalable, allowing users to easily add or remove nodes and workloads as needed. However, scaling up a K8s cluster dramatically increases the complexity of the system and the amount of data it generates. Despite this added complexity, users' requirements and expectations for performance and reliability do not change. As a result, the need for efficient monitoring and diagnosis of issues in K8s clusters has become increasingly important [1].

K8s packs a rich set of self-monitoring capabilities, of which the most powerful one is the *Auditing* feature [2]. K8s audit logs provide a detailed, chronological record of all the activity in a cluster, performed by all actors, including users, applications, and system components. This information is invaluable for diagnosing issues, understanding the system's behavior, and ensuring compliance with security and regulatory requirements. However, auditing in K8s does not come free of any challenge.

When a user performs any action on a K8s cluster, what may look like a simple request is translated under the hood into a series of actions that are performed by multiple components. Creating a Pod may involve the creation of Volumes, the scheduling of the Pod on a Node, the creation of a Service, and so on. This is normal and expected behavior, since the cluster must manage multiple interdependent components dynamically and must adapt to changes in real-time. However, from a monitoring perspective, each of these actions generates its own log line, complicating the analysis of logs and the understanding of the system's state.

In audit logs, all the lines *correlated* to a single user action are usually scattered, and, as the interactions increase in complexity and length, this problem becomes even more pronounced. Furthermore, correlating these lines by hand is at best a daunting task. Each line contains dozens and dozens of fields that may or may not be relevant, and there is often no direct way of linking lines together. To further complicate the matter, even small-sized clusters with few nodes are inherently verbose. For example, nodes need to constantly report their status, even when no

* Corresponding author.
*E-mail addresses:* matteo.franzil@unitn.it (M. Franzil), varmani@fbk.eu (V. Armani), l.diasknob@fbk.eu (L.A. Dias Knob), domenico.siracusa@unitn.it (D. Siracusa).

```
{"kind":"Event","apiVersion":"audit.k8s.io/v1","level":"RequestResponse","auditID":"6f25b082-8730-4bce-a89e-55b7e2d447e5","stage":"ResponseComplete","re
questURI":"/apis/apps/v1/namespaces/rising/deployments?fieldManager=kubectl-client-side-apply&fieldValidation=Strict","verb":"create","user":{"username"
:"mfranzil","groups":["system:authenticated"]},"sourceIPs":["192.168.42.228"],"userAgent":"kubectl/v1.30.2 (darwin/arm64) kubernetes/3968350","objectRef
":{"resource":"deployments","namespace":"rising","name":"test-deployment","apiGroup":"apps","apiVersion":"v1"},"responseStatus":{"metadata":{},"code":20
1},"requestObject":{"kind":"Deployment","apiVersion":"apps/v1","metadata":{"name":"test-deployment","namespace":"rising","creationTimestamp":null,"label
s":{"app":"test"},"annotations":{"kubectl.kubernetes.io/last-applied-configuration":"{\"apiVersion\":\"apps/v1\",\"kind\":\"Deployment\",\"metadata\":{\
"annotations\":{},\"labels\":{\"app\":\"test\"},\"name\":\"test-deployment\",\"namespace\":\"rising\"},\"spec\":{\"replicas\":2,\"selector\":{\"matchLab
els\":{\"app\":\"test\"}},\"template\":{\"metadata\":{\"labels\":{\"app\":\"test\"}},\"spec\":{\"containers\":[{\"image\":\"nginx:latest\",\"name\":\"ng
inx\",\"ports\":[{\"containerPort\":80}]}]}}}}\n"}},"spec":{"replicas":2,"selector":{"matchLabels":{"app":"test"}},"template":{"metadata":{"creationTime
stamp":null,"labels":{"app":"test"}},"spec":{"containers":[{"name":"nginx","image":"nginx:latest","ports":[{"containerPort":80,"protocol":"TCP"}],"resou
rces":{},"terminationMessagePath":"/dev/termination-log","terminationMessagePolicy":"File","imagePullPolicy":"Always"}],"restartPolicy":"Always","termin
ationGracePeriodSeconds":30,"dnsPolicy":"ClusterFirst","securityContext":{},"schedulerName":"default-scheduler"}},"strategy":{"type":"RollingUpdate","ro
llingUpdate":{"maxUnavailable":"25%","maxSurge":"25%"}},"revisionHistoryLimit":10,"progressDeadlineSeconds":600},"status":{}},"responseObject":{"kind":"
Deployment","apiVersion":"apps/v1","metadata":{"name":"test-deployment","namespace":"rising","uid":"d11a8487-b26b-4672-9925-2017b9f9a46a","resourceVersi
on":"6895808","generation":1,"creationTimestamp":"2024-07-03T12:19:13Z","labels":{"app":"test"},"annotations":{"kubectl.kubernetes.io/last-applied-confi
guration":"{\"apiVersion\":\"apps/v1\",\"kind\":\"Deployment\",\"metadata\":{\"annotations\":{},\"labels\":{\"app\":\"test\"},\"name\":\"test-deployment
\",\"namespace\":\"rising\"},\"spec\":{\"replicas\":2,\"selector\":{\"matchLabels\":{\"app\":\"test\"}},\"template\":{\"metadata\":{\"labels\":{\"app\":
\"test\"}},\"spec\":{\"containers\":[{\"image\":\"nginx:latest\",\"name\":\"nginx\",\"ports\":[{\"containerPort\":80}]}]}}}}\n"},"managedFields":[{"mana
ger":"kubectl-client-side-apply","operation":"Update","apiVersion":"apps/v1","time":"2024-07-03T12:19:13Z","fieldsType":"FieldsV1","fieldsV1":{"f:metada
ta":{"f:annotations":{".":{},"f:kubectl.kubernetes.io/last-applied-configuration":{}},"f:labels":{".":{},"f:app":{}}},"f:spec":{"f:progressDeadlineSecon
ds":{},"f:replicas":{},"f:revisionHistoryLimit":{},"f:selector":{},"f:strategy":{"f:rollingUpdate":{".":{},"f:maxSurge":{},"f:maxUnavailable":{}},"f:typ
e":{}},"f:template":{"f:metadata":{"f:labels":{".":{},"f:app":{}}},"f:spec":{"f:containers":{"k:{\"name\":\"nginx\"}":{".":{},"f:image":{},"f:imagePullP
olicy":{},"f:name":{},"f:ports":{".":{},"k:{\"containerPort\":80,\"protocol\":\"TCP\"}":{".":{},"f:containerPort":{},"f:protocol":{}}},"f:resources":{},
"f:terminationMessagePath":{},"f:terminationMessagePolicy":{}}},"f:dnsPolicy":{},"f:restartPolicy":{},"f:schedulerName":{},"f:securityContext":{},"f:ter
minationGracePeriodSeconds":{}}}}}}]},"spec":{"replicas":2,"selector":{"matchLabels":{"app":"test"}},"template":{"metadata":{"creationTimestamp":null,"l
abels":{"app":"test"}},"spec":{"containers":[{"name":"nginx","image":"nginx:latest","ports":[{"containerPort":80,"protocol":"TCP"}],"resources":{},"term
inationMessagePath":"/dev/termination-log","terminationMessagePolicy":"File","imagePullPolicy":"Always"}],"restartPolicy":"Always","terminationGracePeri
odSeconds":30,"dnsPolicy":"ClusterFirst","securityContext":{},"schedulerName":"default-scheduler"}},"strategy":{"type":"RollingUpdate","rollingUpdate":{
"maxUnavailable":"25%","maxSurge":"25%"}},"revisionHistoryLimit":10,"progressDeadlineSeconds":600},"status":{}},"requestReceivedTimestamp":"2024-07-03T1
2:19:13.573094Z","stageTimestamp":"2024-07-03T12:19:13.587796Z","annotations":{"authorization.k8s.io/decision":"allow","authorization.k8s.io/reason":"RB
AC: allowed by ClusterRoleBinding \"mfranzil-cluster-admin-binding\" of ClusterRole \"cluster-admin\" to User \"mfranzil\""},"label":65936}
```

**Fig. 1.** A screenshot of a K8s audit log line of a Deployment action.

Pods are running on them. As workloads are added and the cluster scales up, the amount of generated data exacerbates the correlation problem and also becomes a challenge in terms of storage and processing.

Nothing makes a point clearer than a concrete example: in a controlled experiment using a three-node K8s cluster (details follow in later sections), we observed a considerable level of complexity in the system's behavior. We ran 238 distinct K8s actions, executed a total of 1155 times. Among these, one of the most common, and practically unavoidable, was the creation of a Deployment: a K8s object responsible for declaratively managing replicated pods, including automated rollouts and rollbacks. Every time this action was triggered (what we refer to as the *triggering* event), K8s audit produced a bulky log line, like the one shown in Fig. 1, containing 4700 characters. In addition to that, it produced, on average, another 16 similarly large lines, describing secondary actions related to it. These additional events are often neither directly linked together nor uniquely associated with the Deployment action alone; indeed, many of them can be triggered by different actions across the system. In our contained experimental setting, Deployment actions occurred 40 times and resulted in 636 log lines. If that does not already suggest a tangled puzzle, consider this: deleting a Namespace generates even more complexity, as each resource within the Namespace must first be assessed and then individually removed. In just three Namespace deletions, our system produced 408 log lines, all unrelated to the original triggering event.

Auditing was created specifically to address fundamental questions on cluster actions: what happened, when, who did them, and so on [2]. Yet, we believe that, used as it is, it fails at its purpose. Analyzing data line by line, like most current tools do [3,4], is inefficient and does not leverage the potential of the information available.

To address these shortcomings, in this paper, we propose a novel approach to the analysis of K8s audit logs, called **K8NTEXT**. Our idea is based on the reconstruction of *contexts*. A context is defined as the action that an actor performs on the cluster, plus all the supplementary events that were directly caused by that action. Using a Deep Learning (DL) model, we group correlated lines together, reducing noise in the log and allowing users and machines alike to focus on the most relevant information while retaining all information produced by the auditing system. We evaluate K8NTEXT with regard to its performance, scalability, and expressiveness, and show that K8NTEXT both accurately clusters related actions together and does so in a scalable and efficient manner.

We believe that by structuring audit data into meaningful contexts, our approach lays the necessary groundwork for enabling real-time analysis, an essential step toward timely detection of misconfigurations, anomalies, and security threats in K8s environments.

The rest of the paper is organized as follows. Section 2 provides a brief overview of the background information that is necessary to understand the rest of the paper. Section 3 presents the motivation behind our work, while Section 4 discusses related work in the field of log analysis and correlation. Section 5 describes the design and implementation of K8NTEXT, including the architecture of the DL model and the clustering algorithm. Afterwards, Section 6 extensively evaluates K8NTEXT, focusing on the capabilities of its DL model and the performance of the clustering algorithm. Finally, Section 8 concludes the paper and discusses future work.

## 2. Background

In this section, we present a small review of K8s and its audit logging feature.

### 2.1. Kubernetes

Kubernetes (K8s) [5] is an open-source container orchestration platform that automates the deployment, scaling, and management of containerized applications. It was originally developed by Google, and it is now maintained by the Cloud Native Computing Foundation (CNCF). K8s is designed to be extensible and scalable, allowing users to deploy and manage containerized applications across a cluster of machines.

Fig. 2 shows a simplified view of the architecture of a K8s deployment. At a high level, K8s is composed of a control plane and a set of worker nodes. The former is responsible for managing the cluster, scheduling workloads, and ensuring that the state of the cluster converges towards the state desired by the user. Worker nodes, on the other hand, are responsible for running the actual workloads – in the form of containers – and reporting back to the control plane their status. The control plane and worker nodes communicate in an asynchronous manner, using a well-defined set of APIs that are exposed by the K8s API server.

K8s is designed to be highly scalable. Worker nodes can be added or removed from the cluster dynamically, and the control plane can be replicated to provide high availability. Workloads themselves can be replicated arbitrarily, and K8s will ensure that the desired number of replicas are running at all times. This is achieved through the use of controllers, which are responsible for managing the lifecycle of a given resource. Finally, a rich set of authentication and authorization mechanisms enables Access Control (AC) to the cluster and allows multiple users to interact with it.

### 2.2. Audit logging

K8s provides an *audit logging* feature that allows users to track all API activity that takes place within the cluster [2]. *Everything* that happens in a cluster is logged: from users authenticating and creating Pods, to controllers creating new resources, to the API server itself modifying the state of the cluster.

**Fig. 2.** Simplified overview of the interactions in a K8s cluster. Yellow components are the possible outputs of the audit logging feature.

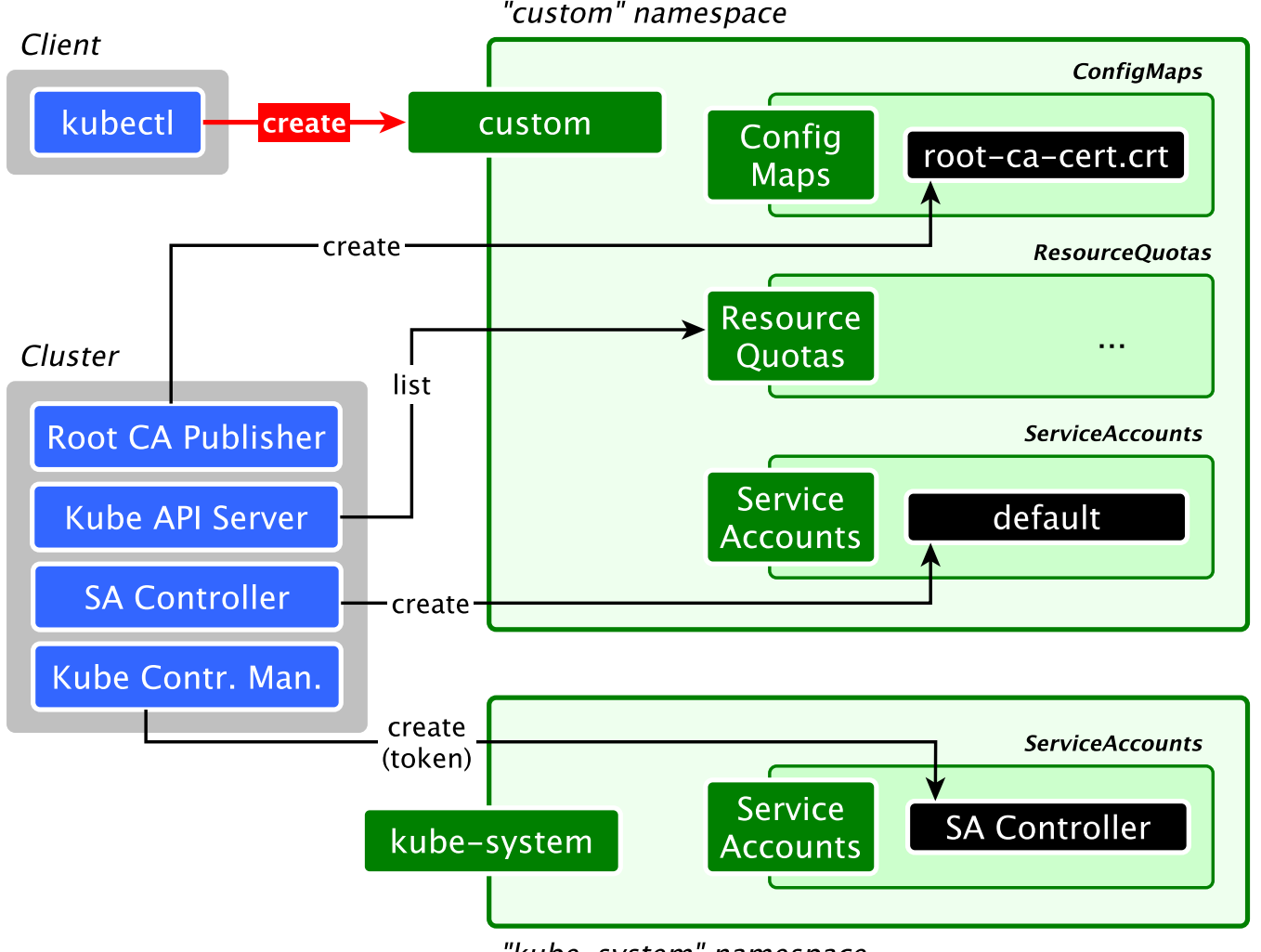

**Fig. 3.** Interactions between components when creating a namespace. On the left side, the various user agents are shown. Each outer box represents a namespace, while inner boxes represent resource types and dark boxes represent objects. Arrows represent interactions between components, with the verb used in the action.

Each API request is logged in a structured format (JSON) and its content can be configured with different levels of granularity. First, cluster administrators define an *audit policy* that specifies the required level of detail for the audit logs. As an example, policies may choose to save only the metadata of the request, or the entire request and response bodies. Furthermore, administrators can specify which resources should be audited and which should be excluded. Finally, the logs can be sent to either a file or a webhook (the yellow boxes in Fig. 2).

## 3. Motivation

Audit logs are a crucial component of any system that aims to be secure and reliable, and in K8s, they are no exception. However, they have so far struggled to strike mainstream adoption [6,7]. This can be attributed to a twofold issue: the sheer amount of data they generate and the complexity of extracting data from them. In this section, we present the motivation behind this work, discussing the importance and challenges of these issues.

### 3.1. Contextualization

The first and foremost issue when dealing with K8s audit logs is the lack of a *context* tying different log lines together. As mentioned above, *any* action – e.g., an API call by a user – that happens in the cluster is logged. However, executing one of these actions will usually result in the cluster performing one or more additional actions as a direct or indirect consequence, essentially causing a cascade effect.

Let us consider, for instance, a simple scenario in which a user creates a namespace. On the client side, all that the end user sees after typing `kubectl create [namespace]` is a simple confirmation message. However, on the server side, several actions are performed, both sequentially and in parallel, by different actors. After generating the *Namespace* object, the Kube Controller Manager will provide the Service Account controller with a token to access it. Then, the Service Account controller will generate the `default` service account in the new namespace. In the meantime, the Root Certificate Authority (CA) Publisher will generate a new certificate for the new namespace, and the Kubelet will start watching for changes to it. All of these actions are performed in parallel, and while none of them directly involve the user who created the namespace, they were *triggered* by the user's action. Fig. 3 illustrates this scenario.

This lack of context is immediately apparent even when giving a cursory look at an audit log. This namespace creation could be scattered across twenty lines in a small cluster, but across hundreds in a busy, large cluster. To make matters worse, each line is a stand-alone JSON that could be by itself large: in this scenario, each line contains 3450 characters on average, or a number of fields varying between 40 to more than 100.

In this vast sea of information, it is easy to lose track of the original action that triggered all the others. The original, *triggering* action is logged as a single event, with its request and response payload, clearly identifying the user who created the namespace. However, the actions that are performed as a consequence of the original action are logged as separate events, with their own request and response payloads, potentially spread dozens of lines apart from each other. In fact, given the distributed nature of clusters, such events may be logged almost simultaneously or in a few seconds, depending on the cluster's size and network health.

On the surface, this might seem like a minor and dismissible issue, as the core action should always be logged correctly. However, as we show below, disregarding the context of an action can lead to false alarms, misidentification of issues, and, in the worst case, security breaches. Furthermore, this "modus operandi" of K8s, where different components in parallel perform actions, is not limited to namespace creation, but is pervasive throughout the system. Deployments get scaled up and down, Services get exposed, and Pods get created and destroyed, all in parallel and in response to different actions with little to no user intervention and, most importantly, intertwined with each other without clear separation.

### 3.2. Verbosity and complexity

To exacerbate the contextualization problem, K8s audit logs are inherently verbose and complex. Even in a small, freshly configured clus-ter, enabling audit logging right out-of-the-box will generate a hefty flow of information. Indeed, as K8s is by design a distributed system, the number of components that interact with the API server and with each other is large, and each and every single small interaction is logged, when no filter is applied. As a non-exhaustive list, these “background” interactions include: heartbeats[1] between the control plane and worker nodes, service account token requests and renewals, control plane components watching for changes in the cluster state, and so on. All of these interactions are logged, and some of them could be confused with the results of user actions.

Furthermore, as the cluster scales up, the amount of generated data increases dramatically. Multiple users and actors may interact with several resources at once, complicating the context reconstruction process. An innocuous-looking action such as a Deployment creation could generate ten lines of logs in a small cluster. In a bigger cluster with several nodes, the same action could generate hundreds of lines, as the Deployment controller will create several Pods, which in turn will generate several events, and so on. Indeed, the number of lines generated by a single action is not fixed, but rather depends on the cluster’s size and complexity.

To further complicate the matter, these lines are hard to contextualize and filter and extremely varied. First, the K8s API is vast and complex [5], resulting in hundreds or thousands of distinct behaviors that may or may not be relevant to the task at hand. Second, even with the efforts made by the API Machinery Special Interest Group (SIG),[2] by the nature of K8s code, distinct SIGs will create objects in different ways, generating JSON with fields that are not entirely compatible. At last, the API presents some corner cases that are not immediate to address. For example, the `objectRef` field in the logs refers to the object being acted upon, but it is often unclear what the object is, as the field is not always populated. Such a scenario makes it incredibly difficult for a human to manually parse the logs – even if already familiar with the K8s API – and for a machine to efficiently categorize and identify information.

One might argue that a correct configuration, verbosity and an efficient log management system can alleviate these issues. Yet, deploying effective configurations is not always straightforward, especially in large, complex clusters. Knowing in advance which data types should monitor and which not is often unfeasible, as K8s’ highly extensible and dynamic API allows for the creation of custom resource which may not be known in advance. On the other hand, using a log management system such as Elasticsearch [8] or Splunk [9] can help in filtering and querying the logs, but it does not solve the problem of the complexity of the logs themselves. Indeed, these systems are designed to store and query logs, not to understand them.

### 3.3. Motivating examples

Finally, we present three real-world scenarios that illustrate the importance of audit logs in K8s, and how the lack of context and the verbosity and complexity of the logs can hinder their usefulness.

#### 3.3.1. Malfunctioning components

A system administrator is in charge of a large K8s cluster with multiple nodes, namespaces, and users. In such a complex system, issues are bound to happen: nodes run out of space, Pods crash, and components fail. The administrator needs to identify the root cause of the issue quickly: to do so, they employ a system of log collection and alerts to notify them of any issues. However, linking the alerts to the actual cause of the issue is often difficult.

Every ten seconds, nodes exchange a heartbeat message with the control plane. If a Node stops sending heartbeats or sends malformed ones, the control plane marks the Node as unhealthy and starts the eviction process. This process may generate a lot of logs, which are hard to correlate, contextualize, and filter. On the other side of the spectrum, problems affecting the underlying operating system (e.g., running out of disk space, the Container Runtime Interface (CRI) crashing, etc.) may not generate any logs at all. In both cases, the administrator is left with a lot of logs to sift through and no clear way to identify the root cause of the issue.

As an example, following a successful upgrade of a K8s cluster, the control plane failed to establish secure communication with its nodes. Although the nodes themselves remained healthy–continuing to send heartbeats, the leader election process repeatedly timed out, and the control plane could not schedule or deploy any new workloads. Inspection of the extracted log entries revealed that every attempt by the kube-controller-manager to fetch its own Lease object under the `coordination.k8s.io` API was being met with HTTP 401 "Unauthorized" responses. This kind of problem is not easy to identify, and even during a cluster’s boot-up phase, the audit logging module will generate thousands of lines. This breakdown in TLS certificate validation effectively severed the control plane-node channel and halted cluster operations.

#### 3.3.2. False alarms

In large multi-tenant K8s clusters, multiple users frequently deploy large Helm charts for testing purposes. These deployments involve legitimate actions like creating namespaces, service accounts, and secrets, which are difficult to distinguish from malicious activities, but may generate many alerts. The security team needs a way to quickly identify the context of these actions and filter out the noise generated by the components of the control plane.

As outlined in this paper, deploying K8s objects generates a lot of logs, which are hard to correlate, contextualize, and filter. Tools such as Falco [3,10], while proficient in parsing the logs, are only able to detect incidents line by line and do not perform any temporal correlation. For example, deploying a Helm chart for Prometheus[3] generates at least 1300 lines of audit log. Feeding these 1300 lines into Falco outputs 389 lines of Falco logs. Most of these errors and warnings relate to new control plane components being created and interacting with secrets, for example:

```
1  Error K8s Secret Get Successfully (user=system:
2  serviceaccount:rising:kube-[..]-operator
3  secret=prometheus-kube-[..]-prometheus-web-config
4  ns=rising resource=secrets resp=200 decision=allow
5  reason=RBAC: allowed by ClusterRoleBinding
6  "kube-[..]-operator" of ClusterRole
7  "kube-[..]-operator" to ServiceAccount
8  "kube-[..]-operator/rising")
```

None of these are malicious (provided the request was legitimate in the first place).

#### 3.3.3. Evading detection

The department of a company that is in charge of security has just received a report that a user has been compromised. The hacker knows how to avoid detection, for example by leveraging proxy forwarding, keeping persistent connections open, or by abusing legitimate commands to extract information from the cluster. The security team is unable to detect the breach, and the hacker is able to exfiltrate a large amount of data from the cluster.

Some corner cases of the K8s API are extremely tricky to identify and are often abused or misconfigured. Some of these behaviors can be detected line-per-line, such as `kubectl exec`. However, as mentioned in the previous section, such types of actions will likely generate several false alarms by systems such as Falco. On the other hand, the use of Node and Pod proxies is not detected by tools such as Falco at all. In both cases, an attacker could easily evade detection by opening a persistent connection to the cluster and exfiltrating data slowly over time.

[1] Heartbeats are periodic API exchanges between the control plane and worker nodes to ensure that the nodes are still alive and healthy.

[2] SIGs are working groups within the K8s community that focus on specific areas of the project.

[3] https://artifacthub.io/packages/helm/prometheus-community/kube-prometheus-stack.

Assume the attacker has successfully leaked a user's credentials in the cluster. Even if a security expert suspects that a user has been compromised, looking specifically at the audit log is not straightforward since it only shows individual API calls without enough context to tell routine admin work from malicious activity. On top of that, attackers often stick to valid API calls or built-in escalation paths that raise no errors. Hence, the administrator needs to manually piece together behavioral patterns, network indicators, and threat intelligence just to hint at the real compromise path.

## 4. Related work

This section provides an overview of the related work. First, we focus on the state of the art in K8s audit logging, then we provide an overview of other log analysis approaches, including classic log analysis and provenance. Finally, we discuss the use of ML in log analysis.

### 4.1. Cloud and Kubernetes audit logging

The concept of audit logging in cloud platforms is relatively new. Amazon Web Services (AWS) introduced its CloudTrail service [11] in 2013, which provides a record of all actions performed on an AWS account. Google introduced its Cloud Audit Logs [12] in 2017, which provides a similar service for Google Cloud Platform (GCP) resources. Finally, K8s audit logging [2] was introduced in 2017, with the release of version 1.7.[4]

Unfortunately, K8s audit logging has so far garnered little attention from the research community. To the best of our knowledge, PerfSPEC [13] is the first and only publication that actively employs K8s audit logs, albeit as a means of proactively identifying overheads when managing security policies.

On the industry side, Sysdig [14,15] and Falco [3] are two companies that have developed tools to analyze K8s audit logs. Sysdig is a cloud-native security platform that provides a wide range of features for monitoring and securing cloud-native applications. It can ingest audit logs from a K8s cluster and provide insights into the security of the cluster. Audit logs are shown in Sysdig as a list of events, with the ability to filter and search for them. However, Sysdig does not perform any correlation between different log lines nor does it provide a comprehensive analysis of the relationships between events in the audit logs. Falco, on the other hand, focuses on real-time security monitoring and can detect anomalous behavior based on audit logs. Falco ships with a set of pre-defined rules that can be used to detect suspicious activity in a K8s cluster. These rules are customizable and can be tailored to the specific needs of the user. However, they are very prone to false positives, as they are based on a set of policies that may not be applicable to all environments. Finally, another tool that recently started employing K8s audit logs is Lacework [4], a cloud security platform that provides insights into the security of several cloud services. Lacework can, too, ingest audit logs from a K8s cluster, although again it performs no correlation between different log lines, and the extent of the anomaly detection capabilities is undisclosed.

### 4.2. Unstructured log analysis

The field of log analysis is vast and has seen a surge in popularity in recent years. Both academia and industry have been actively researching the topic, with a plethora of papers exploring the topic from both an observability and a security standpoint and using both ML and traditional techniques.

The work described in this paper deals with *audit logs*, which are provided in a structured format. However, the majority of the research in the field of log analysis has focused on *classic* log analysis, i.e., the analysis of unstructured logs written in plain text, such as those from Linux's auditd [16]. Most systems have been using and still use this format, and it comes with its own set of challenges.

To begin with, having an unstructured log means that information must be manually extracted from the log lines, which is a tedious and error-prone task. Works such as Drain [17] use a tree-based structure to parse logs, while others like Spell [18] instead employ longest common subsequence and prefix trees. All of these works focus on the problem of *pattern mining*, i.e., the identification of common patterns in log lines.

Once these common patterns are identified, log lines must be consistently clustered together. In classic logs such as *system logs*, linking information includes the Process Identifier (PID), the files being accessed, or the network connections being made. Several works first parse the logs into *log templates*, then link lines together by means of the information found in each line. For example, LogCluster [19] vectorizes logs and then attempts to find past log lines that are similar to the current one. On the other hand, LogMiner [20] clusters system logs into *behaviors* (e.g., deleting files, compressing, etc.) by analyzing system calls made in the system. All of these works, however, are of little use in the context of K8s audit logs, since JSON logs are already structured and easily parsable.

### 4.3. Provenance

Strictly correlated to log analysis is the concept of *provenance*. In security, provenance refers to the reconstruction and analysis of the history of a system, in order to identify the root cause and the sequence of events that led to a security incident [21]. In this context, logs are a vital, almost irreplaceable resource for security experts, both in classic environments and in cloud-native ones [22,23].

Provenance has been extensively studied in the literature, with dozens of works showing how information can be efficiently extracted from logs and used to reconstruct the history of an intrusion, or to identify the root cause of a failure. Core to most works is the concept of *causal dependency*, which refers to the relationships between different events in a system. For example, if event A is the cause of event B, then A is said to be a *causal predecessor* of B.

Works in the field that perform causal analysis include NoDoze [24], ProvGRP [25], HERCULE [26], DEPCOMM [27]. Other works that improve on this approach include Winnower [28], which proposes a system that uses grammatical inference over the single logs, clusters them, and generates behaviors using a state machine; and ProvG-Searcher [29], which performs embedding over the logs and then uses a graph-based approach to identify the relationships between them. Finally, LogApprox [30] quantifies how much redundancy in logs can be removed while still being able to reconstruct the original log lines.

Overall, these works perform *post-mortem analysis*, i.e., they analyze logs after they have been generated and stored elsewhere. Post-mortem analysis is powerful and can provide fine-grained intelligence. However, when dealing with systems that constantly generate large amounts of logs, timely analyzing them becomes unfeasible, as does storing all the logs for later analysis. To address this issue, some works have proposed *real-time analysis* of logs, i.e., the ability to analyze logs as they are generated. In the literature, works such as HOLMES [31], Sleuth [32], and NodeMerge [33] all perform real-time analysis of logs. However, these works are still limited to classic logs and do not take into account the specificities of distributed environments. When multiple components possibly scattered across multiple machines are involved, the relationships between events can become complex and difficult to identify. Aiming to untangle such complexity, works such as CLARION [23] and ALASTOR [22] extend the concept of provenance to microservice deployment. Using a combination of data aggregation, resource renaming, and summarization, these works efficiently collect data from workloads scattered across multiple machines and aid in the detection of intrusion. This focus on microservices is not limited to provenance works: the research community is actively exploring microservice tracing for performance monitoring and debugging [34,35]. However, all of these works focus on the analysis of logs generated by the microservices themselves, and do not take into account the audit logs generated by the orchestration layer.

[4] https://kubernetes.io/blog/2017/06/kubernetes-1-7-security-hardening-stateful-application-extensibility-updates/.

### 4.4. Machine learning and log analysis

ML has seen a surge in popularity in recent years, thanks to the ever-increasing amount of data that is being generated by modern systems and the commonplace availability of powerful computing resources. ML models are proficient in quickly identifying patterns and anomalies in logs, either by generalizing from a small set of labeled data (*supervised learning*) or independently identifying patterns by themselves (*unsupervised learning*). As an example, DyCom [36] groups logs into communities and then uses an embedding-based graph encoder to identify the relationships between them. LogGraph [37] employes Graph Neural Networks to obtain a similar result.

With the advent of Large Language Models (LLMs), the field of log analysis has also seen a shift towards using bigger models to analyze and understand the relationships between different events. LLMs are trained on large amounts of text data and can generate human- and machine-like text. They have been shown to be effective in a variety of natural language processing tasks, such as text classification and machine translation. LogPrécis [38] shows promising results in log classification, using a small set of labeled data to fine-tune a pre-trained LLM.

### 4.5. Summary

The state of the art in K8s audit logging is still in its infancy, with little research being done on the topic. The few (mostly non-academic) works that do exist do not perform any correlation between different log lines, limiting their usefulness. On the other hand, existing techniques are ill-equipped to analyze K8s audit logs. Casual analysis solutions are often tailored to system audit logs, whose features such as process IDs and file descriptors aid in establishing relationships between events. Sequential analysis solutions, on the other hand, are often limited to analyzing logs line by line, failing to capture any dependency at all. Finally, rule-based approaches require creating and maintaining a vast, ever-growing catalog of event patterns to correctly contextualize logs. This approach is not only labor-intensive but inherently brittle, as it needs frequent updates to accommodate the constant changes to the K8s API and the introduction of new features.

Thus, there is a need for a new approach to analyze K8s audit logs, one that can efficiently capture the relationships between events and provide a comprehensive analysis of the logs while doing so efficiently, in real-time, and without requiring a vast catalog of rules. ML is a promising approach to achieve this, as it can quickly identify patterns and anomalies in logs and can be trained to understand the quirks of K8s audit logs. However, some specific challenges such as linking complex relationships between events, handling a high volume of logs and fields, and supporting a vast number of possible events, have not yet been addressed in the literature, leaving a gap that needs to be filled.

## 5. K8NTEXT

To address the limitations identified in Section 3 and unresolved by existing tools, we present K8NTEXT. K8NTEXT is a pipeline that leverages machine learning to automatically identify relationships between log lines, enabling effective event contextualization. Our approach transforms the raw, unstructured audit logs into structured, queryable contexts that provide a holistic view of cluster activities.

### 5.1. System architecture

Fig. 4 shows the system architecture of K8NTEXT. The system is designed to be modular and extensible, allowing users to customize it to their needs. The main components of the system are:

- **Log pre-processing**: audit logs coming from the K8s cluster are parsed, filtered and reordered;
- **Label prediction**: a ML model predicts the label of the log lines;
- **Context reconstruction**: the log lines are clustered into contexts using the predicted labels.

K8NTEXT exposes the now-clustered log lines in various formats, optionally including a query engine that allows users to query the contexts that have been reconstructed and visualize them in a user-friendly way. The query engine is not mandatory, and users can choose to use K8NTEXT as a simple log parser that outputs the clustered log lines in a format of their choice. K8NTEXT is very fast and can perform inference in milliseconds even on large clusters, enabling real-time consumption of audit logs.

All components of K8NTEXT are written in Python 3.10 and use the Keras 3.0 library [39] for the ML model, making it agnostic to the underlying ML framework. The query language is implemented using Lark [40], a Python library for parsing context-free grammars. The source code of K8NTEXT is available on GitHub [41].

### 5.2. Contextualization

We first explain the concept of *contextualization* and how it is implemented in K8NTEXT. The goal of K8NTEXT is to reconstruct the context of an event, i.e., the set of events that are related to it. An **event** $e$ is defined as a single log line, while its **context** of an event $e$ as the set of events $E$ s.t.,

1. $\exists \hat{e} \in E : \hat{e}$ is the **triggering event** or the event that directly caused all the other events in $E$ to happen,
2. $\forall e \in E, e \neq \hat{e}$, $e$ happened as a *direct* or *indirect* consequence of $\hat{e}$,
3. the timespan $(t_0, t_1)$ that contains all the events in $E$ is as small as possible.

K8s is at its core a distributed system. Thus, events may happen in parallel or in quick succession. As a result:

- the triggering event $\hat{e}$ may not be the first event in $E$,
- the events in $E$ may be unordered,
- the timespan $(t_0, t_1)$ can contain events that are not directly related to the triggering event,
- the timespan $(t_0, t_1)$ may be excessively large or open on the right.

Furthermore, some tools such as `kubectl` and its *client-side apply*[5] intentionally perform some initial actions (e.g., getting information on the namespace of an object) before actually sending the request to the K8s API, further complicating the task of identifying the triggering event.

Table 1 summarizes the symbols used in this section to define and describe the various components of K8NTEXT, both in this section and in the rest of the paper.

### 5.3. Log parser

To begin with, audit logging must be properly configured in the K8s cluster [2]. K8s auditing natively supports several criteria for filtering events at the source, such as the resource type or the user that performed the action. However, in our implementation, no filtering is performed. This allows K8NTEXT to ingest all events generated by the cluster, and thus, to have a comprehensive view of the cluster's activity and to truly capture every dependency and interaction between events. Filtering at the source would reduce the amount of data ingested, but it would also provide K8NTEXT with an incomplete view of the cluster's activity, potentially leading to missed dependencies and interactions between events.

[5] https://kubernetes.io/docs/reference/using-api/server-side-apply/#comparison-with-client-side-apply.

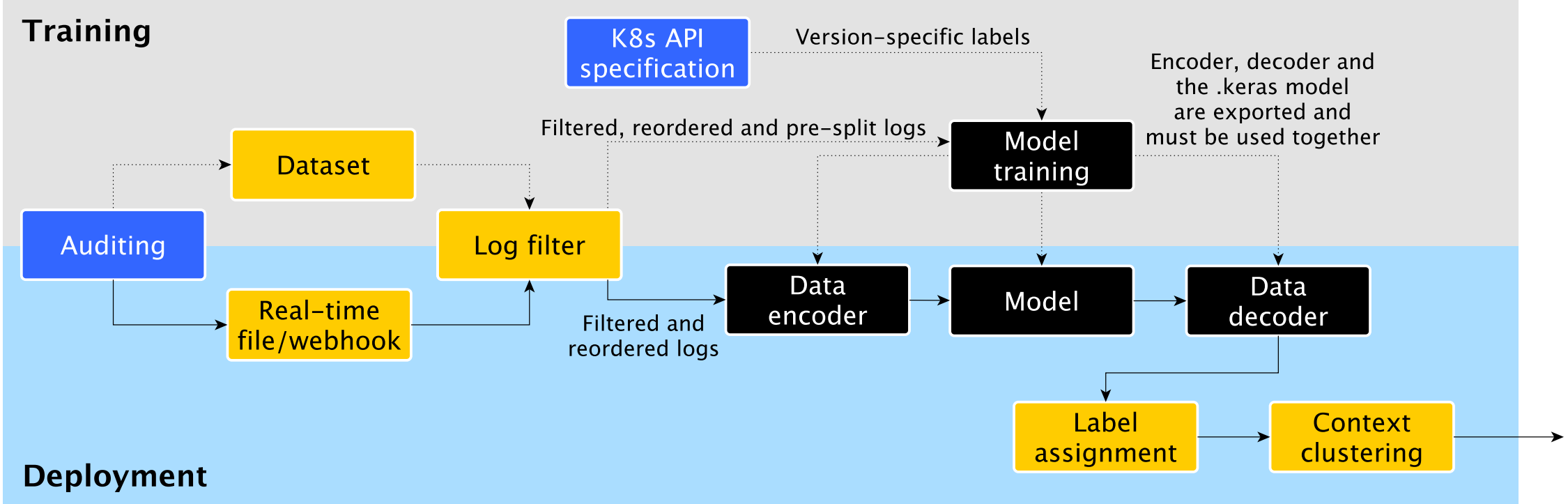

**Fig. 4.** System architecture of K8NTEXT. Dotted lines indicate the flow for the training phase, while solid lines indicate the flow for the inference phase. Black boxes represent the DL components, blue (dark) boxes represent the K8s components, and the yellow (light) boxes represent all other log-handling components. For clarity, the query engine is optional and is left out of the figure. (For interpretation of the references to colour in this figure legend, the reader is referred to the web version of this article.)

**Table 1**
Table of symbols.

| Symbol | Description |
|---|---|
| $e$ | Event |
| $\hat{e}$ | Triggering event |
| $E$ | Set of events |
| $t_0, t_1$ | Timespan containing events in $E$ |
| $\|E\|$ | Number of events in $E$ |
| $W$ | Window size |
| $\|F\|$ | Number of features |
| $\|L\|$ | Length of the label tuple |
| $m$ | Highest integer in the label tuple |
| $n$ | True if the object is namespaced |
| $u$ | True if the action was on a single object |
| $r, s, v$ | Resource, subresource, and verb identifiers |
| $f_A$ | API version, resource and subresource → $(r, s)$ |
| $f_V$ | Verb name → $v$ |

Once the logs are generated, they are sent to the log parser. The log parser reads the logs generated by the K8s cluster and filters, reorders, and normalizes them in a way that is suitable for the label predictor and the cluster's version of the API. `null` values are added for some missing fields (see Section 5.4), and some lines unrelated to API calls are discarded.

For compatibility reasons, the log parser supports any CustomResourceDefinition (CRD) added to the cluster automatically, but without extracting any context. CRDs are by nature dynamic and diverse, a trait hard to capture with a fixed-size model. We leave the extraction of contexts from CRDs as future work.

### 5.4. Fields and features

Each log line is a distinct JSON object that contains a variable set of key-value pairs. Some of these fields are fixed and appear in every log line, while others are optional and may or may not be present depending on the type of event being logged, the status of the cluster, and its configuration. K8s defines in its documentation three types of audit levels: *Metadata*, *Request*, and *RequestResponse*. In turn, we categorize key-value pairs generated by each audit level as **Metadata**, **Request**, or **Response** fields.

Metadata fields are always present in the logs and comprise key information about the event: its timestamp, the user that performed the request, the target of the request, and so on. Request and response payloads, on the other hand, contain a wealth of information that greatly varies even between different events of the same type. Their structure is unpredictable, containing fields that are sometimes duplicate, nested, or in non-string format (such as a raw YAMLs as a value of a key). Sometimes, they may not be present altogether: some API calls do not require a request body, while others do not return a response body.

Table 2 shows a selection of the most relevant features we chose for the model, while the complete list is available in the source code. To obtain the features, each JSON is flattened, i.e., all nested fields are extracted and converted into a single-level dictionary. This allows us to easily access the fields we are interested in, regardless of their nesting level.

Choosing the features was a two-phase process. First, we performed a systematic analysis of the logs generated by several K8s clusters, including both production and test clusters. We thoroughly examined the logs and the K8s source code to understand how the logs are generated and what fields are available in each log line. This allowed us to identify key linking fields across different types of resources in the API. Using this approach, we obtained 50 features. These 50 features were then further reduced to 39 by means of an iterative feature selection process, removing some redundant and less relevant ones.

Finally, in Section 6, the features are evaluated in terms of their impact on the model's performance. We show that while the 39 features could be further reduced, they are sufficient to achieve an optimal accuracy.

### 5.5. Labels

The next step is to define an aggregation mechanism for grouping correlated events together. The goal is to partition a set of log lines into a set of clusters, each containing a triggering event $\hat{e}$ and all the events $E$ that are related to it. Since directly identifying such clusters is complex, we instead introduce an intermediate representation where we aggregate together log lines *related to the same type of action*.

Recall the example of creating a namespace mentioned in Section 3.1. In a cluster, multiple namespaces may be created over time or possibly concurrently. Rather than attempting to uniquely identify each act of creating a namespace plus all the correlated actions, we first group together all the lines that are in some sort connected with the creation of a namespace. To do so, we employ a ML model (described later in detail in Section 5.8) to perform multiclass classification, i.e., we train the model to recognize, for each log line, *the type of action it is*

**Table 2**
Key features used in the model.

| Feature | Description |
|---|---|
| `objectRef.apiGroup` | API group of the object |
| `objectRef.namespace` | If the object is namespaced |
| `objectRef.resource` | Resource acted upon |
| `objectRef.subresource` | Subresource acted upon |
| `user.groups[0/1/2]` | User groups (first three) |
| `userAgent.extra` | Extra information in the user agent |
| `userAgent.tool` | Tool used to interact with the API |
| `userAgent.version` | Version of the tool used |
| `verb` | Verb of the request |
| `requestObject.metadata.ownerReferences.[…]` | Information on the parent object |
| `responseObject.involvedObject.[…]` | Information on additional objects |
| `responseStatus.code` | Status code of the response |

**Table 3**
List of fields used to define the `label` of a triggering event.

| Field | Description | Example |
|---|---|---|
| `objectRef.apiGroup` | The K8s API group of the object | `core`, `apps`, `rbac.authorization.k8s.io` |
| `objectRef.apiVersion` | The K8s API version of the object | `v1`, `v1beta1` |
| `objectRef.resource` | The resource type of the object | `pods`, `namespaces`, `secrets` |
| `objectRef.subResource` | The subresource type of the object | `status`, `scale` |
| `verb` | The action that was performed on the object | `create`, `delete`, `get`, `list`, `update` |

*related to*. To achieve so, we need to define a set of labels that can be used to identify the type of action.

We start from a subset of the features briefly described in the previous section. These features, shown in Table 3, are always present in the logs and do not have any temporal or identity information: they are solely related to the characteristics of the action itself.

Then, we augment this information by introducing two additional fields, `objectRef.namespace` and `objectRef.name`, that are present in the logs only if the object exists in a namespace and the action was performed on a single object, respectively.[6]

Finally, we define two helper functions $f_A$ and $f_V$ that take as input the K8s API version, the resource name, and the subresource, and return a tuple $(r, s)$, where $r, s \in \mathbb{Z}$ are the resource and subresource identifiers, respectively. The function $f_V$ is similar, but it takes the verb name as input and returns a verb identifier. These functions are used to map the K8s API version, resource name, subresource, and verb to a unique identifier. This allows us to define a consistent set of generic behaviors that are common to all K8s clusters running that version.

We can now define the `label` $L$ of a triggering event as a tuple $L = (r, s, v, n, u)$.

- $r, s \in \mathbb{Z}$ are the resource and subresource identifiers as defined by $f_A$,
- $v \in \mathbb{Z}$ is the verb identifier as defined by $f_V$,
- $n \in (0, 1)$ is a boolean that indicates if the object is namespaced $(1)$ or not $(0)$, i.e., if the action has been performed on a namespaced object or on a cluster-wide object,
- $u \in (0, 1)$ is a boolean that indicates if the object is "single", i.e., if the action has been performed on a single object $(1)$ or on multiple objects $(0)$.

To obtain the last two items, we check if the `namespace` and `name` fields are present and not null in the logs. If they are, the object is namespaced and single, respectively. With this definition in place, the **label** of an event is a tuple that uniquely identifies the characteristics of the triggering event of which the log line is part of. Such an approach allows us to maintain a small and manageable set of labels, while still being able to capture the complexity of the K8s API.

Before moving on, we must note that the label is not a unique identifier of the event itself, but rather a unique identifier of the type of action. After the ML model has predicted the label, we obtain clusters, each containing one or more contexts that need to be separated. This process is then performed by the clustering module, described in Section 5.9.

### 5.6. Dataset

Given the lack of publicly available datasets for this task, we generated our own dataset, available along with the source code of K8NTEXT on GitHub [41]. The dataset contains a total of 18,478 log lines collected from a K8s cluster running in a virtualized environment in our premises. To generate this dataset, we created a simple three-node cluster using `kubeadm`. One single node performed control plane duties, while all three nodes could schedule workloads.

Audit logs are agnostic to the number of nodes in the cluster, since they refer to API calls that are handled in the control plane, mostly incoming from itself and users in the system. As a result, we decided to settle on a three-node architecture, a compromise between our computational resources at our disposal and the fact that arbitrarily scaling the cluster would add little value to both the dataset and the evaluation of K8NTEXT.

The dataset was generated in a systematic and reproducible way. Starting from the official API documentation, we generated a list of all the possible endpoints and methods supported by each, then narrowed it down to a list of typical interactions with the API resources.

This list comprises typical interactions with the cluster (such as creation, deletion, inspection of Pods, Deployments, and so on), network management (creation of NetworkPolicies, Ingresses, etc.), storage management (creation of PersistentVolumes, StatefulSets, etc.), and advanced interactions with the cluster (such as Jobs, CronJobs, etc.). We also included interactions with authentication resources, such as Roles and RoleBindings, and the creation and management of users.

We also tried to retain realism in the dataset, simulating failed rollouts of Deployments, concurrent creation of resources, and other edge cases. This increased realism came at a cost of a great imbalance in the dataset. Indeed, resources such as Pods and Deployments are central to a cluster's operation, while others may only see minimal usage. Generat-ing a perfectly balanced dataset would have been both time-consuming and potentially counterproductive. The impact of this choice is investigated in Section 6.5. We also purposely left out some deprecated and lesser-known API resources such as `FlowSchemas`.

[6] K8s differentiates between *namespaced* and *non-namespaced* objects, which can affect how events are processed and interpreted. The latter are cluster-wide objects, while the former are confined to a specific namespace. Some resources may exist in both namespaced and non-namespaced versions. Second, some actions can be performed on a single object, while others can be performed on multiple objects.

Overall, the dataset comprises a total of 238 unique actions out of the possible 1356 that are supported by the K8s API. While this may seem a small number, it must be noted that several combinations are either deprecated or rarely used in practice, and the dataset is not meant to be exhaustive. Our dataset focused on the most common and representative actions, which are sufficient to demonstrate the effectiveness of our approach while providing a flexible model for most use cases.

### 5.7. Label encoding

Using the labels defined in Section 5.5, we manually labeled the dataset, taking care in assigning to each log line a label representing the type of action of its triggering event. To do so, we convert the tuple $L = (r, s, v, n, u)$ into an integer representation. As an example, the act of deleting a ReplicaSet is represented as $L = (r, s, v, n, u) = (17, 0, 2, 0, 1)$, where $r$ is our internal resource ID of ReplicaSet, $s$ is our internal subresource ID (0 in this case), $v$ is our verb ID of `delete`, $n$ is 1 because ReplicaSets are namespaced, and $u$ is 1 because we are deleting a single ReplicaSet. Using a simple and invertible encoding scheme,[7] we convert this tuple into a single integer $l = 70048$. Such a representation is convenient for manual labeling but extremely space inefficient. Indeed, after manually labeling the dataset, we ended up with a total of 1356 unique labels. This means that the model would need to perform multiclass classification over 1356 classes, which is a big yet computationally feasible number.

To improve the model's efficiency, when importing the dataset, we convert the integer back to its tuple representation and then separately one-hot-encode each element of the tuple. Doing so allows us to reduce the size of the one-hot-encoded vector from 1356 to $|L| \times m$, where $|L|$ is the number of elements in the label tuple and $m$ is the maximum integer in all tuples after it has been categorically encoded. In our case, $|L| = 5$ and $m = 59$, so the final size of the one-hot-encoded vector is $5 \times 59 = 295$. This is a significant reduction in size, and it allows the model to be more efficient both in training and in inference.

### 5.8. Deep learning model

Once the log has been parsed, encoded, and normalized, it is sent to the deep learning model for label prediction.

#### 5.8.1. Input and output shapes

Our model was designed from the ground up to be modular and flexible, allowing it to be easily adapted to different datasets and environments. To do so, we rely on some hyperparameters that are set on the fly before the model training. The key hyperparameter is the **window size** $W$: the number of log lines to consider in each batch. This number is strictly related to the *traffic* proper of the cluster itself. Indeed, smaller clusters with less traffic may require smaller window sizes, while busier clusters with several nodes and requests per second will require a larger window size. The experiments with $W$ are shown in Section 6.2.

The other three main hyperparameters are those that directly affect the number of input and output units of the model. The first one is the **number of features** $|F|$, which is manually set during the model configuration and in this case is equal to 39. Thus, the model requires as input a tensor of shape $(W, |F|)$, where $W$ is the window size and $|F|$ is the number of features.

The second and third ones, on the other hand, determine the output shape of the model. As described in Section 5.5, the model is trained to predict a label that is a tuple of integers. This tuple is composed of $|L|$ elements, and each element is an integer that, once encoded, can take a value between 0 and $m - 1$. Our model outputs a tensor of shape $(W, |L|, m)$: in other words, for each log line in the batch ($W$), we can reconstruct the original tuple $L$ by applying a *softmax* activation function to each element $L_i$ of the output tensor.

A study on the impact of these hyperparameters, among others, is presented in Section 6.

#### 5.8.2. Model architecture

We chose simplicity and efficiency as the guiding principles for the design of our model, aiming to create a model that is easy to understand and can be trained in a reasonable amount of time. The model is based on a Bidirectional Long Short-Term Memory (BiLSTM) architecture, which is a type of recurrent neural network (RNN) that is particularly well-suited for sequence prediction tasks. BiLSTM networks are able to capture long-term dependencies in the input data, making them ideal for processing sequences of log lines. Fig. 5 shows the architecture of the model.

The figure shows the layers of the model, which are described in detail below. The model is designed to take as input a tensor of shape $(W, |F|)$, where $W$ is the window size and $|F|$ is the number of features. The output of the model is a tensor of shape $(W, |L|, m)$, where $|L|$ is the number of elements in the label tuple and $m$ is the maximum integer in all tuples after it has been categorically encoded.

The model consists of the following layers:

- An initial input tensor, which depends on the number of features and the window size $(|F|, W)$;
- Two consecutive BiLSTM layers, with decreasing units and separated by a batch normalization layer;
- A dropout layer;
- A time-distributed dense layer that converts the BiLSTM per-timestep predictions to a per-line output;
- A second batch normalization layer;
- A softmax activation function.

Initial experiments with single Long Short-Term Memory (LSTM) layers were not satisfactory, as they showed poor performance with out-of-order or far-apart log lines. We therefore switched to a BiLSTM architecture, which is able to process the input sequence in both directions, from the beginning to the end and vice-versa. We experimented with several layer configurations, including the number of units in the BiLSTM layers, the number of units in the dense layers, the dropout rate, and the number of layers, using Keras' hyperparameter tuning capabilities.[8]

We settled on a decreasing number of units in the BiLSTM layers, with the first layer having $4 \cdot |F|$ units and the second layer having $3 \cdot |F|$ units. This configuration allows the model to scale well with the number of features, while still being able to capture the dependencies in the input sequence. For the dropout layer, we found that a rate of 0.4 was the best at reducing overfitting while maintaining a good performance. The dropout layer is placed after the second BiLSTM layer, to prevent overfitting on the training set.

The model is compiled using the Adam optimizer and the *categorical focal cross-entropy* loss function, which is a variant of the categorical cross-entropy loss function that is more robust to class imbalance. Early stopping, learning rate reduction on plateau, and other techniques are used to prevent overfitting and improve the model's performance. The model is trained for a maximum of 130 epochs. The training process is monitored using the validation loss, and the best model is saved for later use.

The model's final hyperparameters are described in Table 4. Some of them are further studied in Section 6, while others are fixed and not subject to further analysis.

[7] Its implementation is omitted for brevity, but it is available in the source code.

[8] https://keras.io/keras_tuner/.

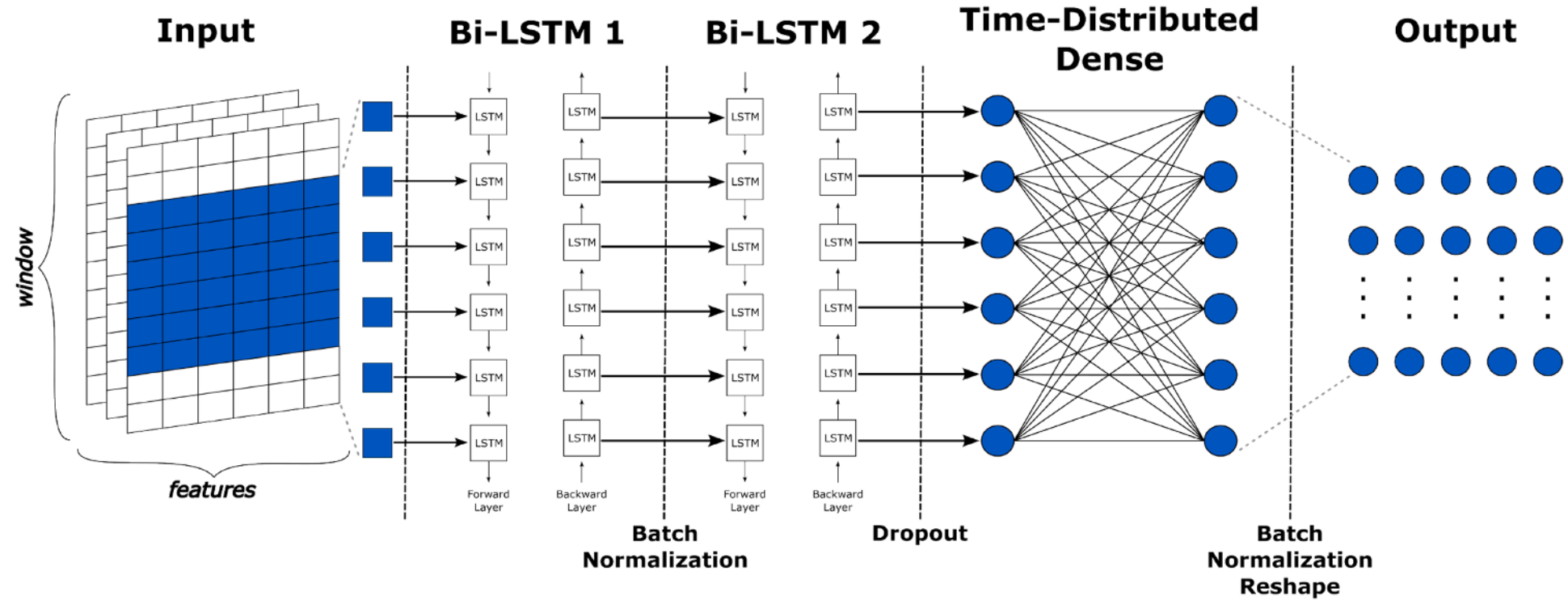

**Fig. 5.** The architecture of the deep learning model used for label prediction.

**Table 4**
Model hyperparameters.

| Hyperparameter | Value |
|---|---|
| Window size $W$ | 60 |
| Number of features $\|F\|$ | 39 |
| Number of labels $\|L\|$ | 5 |
| Maximum integer $m$ | 59 |
| Number of units (1st layer) | $4 \cdot \|F\|$ |
| Number of units (2nd layer) | $3 \cdot \|F\|$ |
| Dropout rate | 0.4 |
| Initial learning rate | 0.004 |
| Early stopping patience | 8 |
| Reduce LR factor | 0.1 |
| Reduce LR patience | 4 |

#### *5.8.3. Batching*

The final issue that must be dealt with before using the model is the batching strategy. With our window size $W$, log lines will appear in up to $W$ different batches. Depending on the position within the batch, the model may predict different labels for the same log line. To obtain a single label for each log line, we implemented a simple majority voting system. For each log line, the model takes the most represented label in the batches it appears in, or the first in case of a tie. This label is the final prediction for the log line and is appended to the log line before being sent to the query engine.

### *5.9. Clustering*

At the end of the labeling process, each log line has been assigned a `label` that uniquely identifies the type of triggering event it is part of. However, we explicitly embedded only a subset of the fields in the `label`, namely, the *generic* fields. This means that while triggering events and their consequences are categorized by type, the specific contexts in which they occur are not yet fully reconstructed. As an example, imagine the creation of two namespaces. Their label will be shared, but the contexts in which they were created will be different. At the moment, we have no way to distinguish between the individual contexts for each namespace creation.

To address this, we need to group log lines that are related to the same specific triggering event instance together. This is a challenging task: log lines may arrive out of order, timestamps may be imprecise due to the distributed nature of K8s, and the linking information between events may be scattered across different fields in the request and response bodies. Moreover, in production environments with high operational frequency and many concurrent users, the clustering algorithm must be robust enough to handle large volumes of log lines without incorrectly assigning events to the wrong contexts.

#### *5.9.1. Windowing strategy*

To ensure scalability and reliability, the clustering algorithm operates on time-bounded windows of log lines rather than processing the entire log stream at once. This approach mirrors real-world scenarios where audit logs are sent in batches to webhook endpoints. K8s uses a default cutoff of 10,000 lines or 30 seconds (whichever comes first) when sending audit events to webhooks.

Our implementation introduces two configurable cutoff based on the aforementioned K8s parameters: a **time threshold** and a set **maximum lines per window**. For each label group identified by the model, we partition the log lines into windows based on these constraints. A new window begins when either the time elapsed since the window's first line exceeds the time threshold or the window reaches the maximum number of lines. This dual-threshold approach prevents both an unbounded growth of clusters and the accumulation of stale data in low-traffic clusters.

Within each window, the clustering algorithm analyzes all lines before making assignment decisions, ensuring that contextual information is properly considered rather than relying on simple sequential assignment strategies. While this windowing approach may introduce errors at the boundaries (e.g., if related events fall into different windows), it provides a practical balance between performance and accuracy, especially in high-throughput environments.

#### *5.9.2. Cluster assignment*

The core of the clustering algorithm is a scoring mechanism that evaluates how well each log line matches existing clusters within the current window. When processing a window, the algorithm first identifies any triggering actions present, i.e., log lines whose label matches their intrinsic characteristics, indicating they directly caused the action rather than being consequences of another event. Each triggering action becomes the seed of a new cluster.

For each remaining log line in the window, the algorithm computes a compatibility score (a simple integer, $s$) with every existing cluster. The score is based on multiple weighted criteria that capture different types of relationships between K8s objects and events. The line is then assigned to the cluster with the highest score, provided it exceeds a minimum threshold ($s = 60$). If no cluster scores above the threshold, a new cluster is created.

The scoring hierarchy can be summarized as follows, in decreasing order of importance:

1. **ObjectRef match**: Direct reference to the same resource (namespace, name, type) is the strongest indicator of relatedness.
2. **UID relationships**: If a log line's object UID matches an `ownerReferences` UID in the cluster's metadata, or vice versa, this indicates a parent-child relationship. For example, a Deployment creates ReplicaSets that create Pods, and the UIDs link these hierarchically.

3. **InvolvedObject match**: Events (e.g., K8s Event resources) reference other objects via `involvedObject`. Matching this field or matching an involved object to a resource in the cluster indicates correlation.
4. **ClaimRef match**: PersistentVolumes and PersistentVolumeClaims are linked via `claimRef`, and matching this field groups related storage events.

Username and verb are also matched, but they are considered secondary and insufficient to warrant clustering on their own. Additionally, the algorithm includes targeted heuristics for known K8s behaviors. For instance:

- When a user creates a resource in a namespace, `system:apiserver` often performs a `list` operation on `limitranges` in that namespace. The algorithm detects this pattern and assigns these system actions to the user's context with a high score (90 points).
- When a namespace is deleted, the system performs `deletecollection` and `get` operations on all resources within that namespace. The algorithm identifies these cascading actions and groups them with the namespace deletion context (95 points).

#### *5.9.3. Control plane and special cases*

Some clusters may benefit from merging all control plane actions (actions performed by system components like the API server and the Kube Controller Manager, etc.) into a single dedicated context. This is controlled by a configurable parameter. When enabled, all control plane actions are assigned a fixed UUID, allowing auditors to separate user-initiated actions from background system operations.

The clustering algorithm also handles edge cases gracefully. For instance, single-line windows are trivially assigned to their own cluster. Windows with no triggering actions distribute lines based on the best scoring matches, creating new clusters as needed. The explicit decision hierarchy ensures that high-confidence matches (e.g., ObjectRef or UID relationships) take precedence over lower-confidence heuristics.

### *5.10. Querying*

Finally, K8NTEXT optionally provides a way to query the contexts that have been reconstructed. The query engine is not mandatory, and users can choose to use K8NTEXT as a simple log parser that outputs the clustered log lines in a format of their choice.

When enabled, users can query the contexts that have been reconstructed using a simple query language. Using classic boolean operators such as $\wedge, \vee, <, \geq, \neg$, and the `exists` and `regexp` operators, users can express complex queries that are then parsed by the engine. For example, the query `verb == "create" and objectRef.namespace == "default"` will return all the contexts that are related to the creation of an object in the `default` namespace.

The power of such a system derives from the fact that users express queries in terms of the fields of the triggering event they desire to investigate, without having to perform complex subqueries to retrieve tailored contexts for each type of event. This contrasts with tools such as `jq`, which, while powerful, cannot differentiate between the triggering event and other events in the context.

## 6. Deep learning model evaluation

This section evaluates K8NTEXT by assessing its prediction model's performance.

All the tests were conducted on an Ubuntu 24.04 LTS virtual machine with 24 GB of RAM, 10 vCPUs based on an Intel Xeon Gold 5218R CPU, and a PCI-passed-through NVIDIA A5000 GPU. Each experiment outlined in this section was run twenty times, and the results were averaged to obtain a more accurate estimate of the model's performance. The source code used to run the experiments is provided alongside K8NTEXT's implementation [41].

**Table 5**
Time and F1 score of the model with different window sizes.

| $W$ | Time (s) | F1 score |
|---|---|---|
| 5 | 177.50 ± 27.45 s | 0.8103 ± 0.0240 |
| 10 | 195.15 ± 42.63 s | 0.9267 ± 0.0173 |
| 20 | 240.10 ± 41.85 s | 0.9681 ± 0.0093 |
| 30 | 238.35 ± 36.24 s | 0.9791 ± 0.0099 |
| 40 | 261.55 ± 51.51 s | 0.9808 ± 0.0068 |
| 50 | 300.70 ± 43.77 s | 0.9866 ± 0.0047 |
| 60 | 319.60 ± 43.94 s | 0.9820 ± 0.0082 |

### *6.1. Metrics*

#### *6.1.1. Storage space*

The storage space, expressed in bytes (and multiples), measures the amount of disk space needed to store both the logs and the model. The latter is relatively less important, as our model will rarely exceed some hundreds of megabytes. The former, however, can be quite large, especially in large-scale deployments, and we evaluate how K8NTEXT can reduce the space needed to store the logs.

#### *6.1.2. Time*

The time metric, measured in seconds, indicates the time needed to perform various operations. The initial *training time* is the time needed to train the model on the datasets, which in our case was done using a GPU. The training time is not negligible, but it is not critical, as training is performed only once. The *inference time*, on the other hand, is the time needed to predict labels. Inference time is fundamental for K8NTEXT's usability. We evaluate how it scales with the dataset size, content, number of features, window size, and test-train-split. Finally, the *querying time* is the time needed to query the logs using K8NTEXT, and we evaluate how this time compares to manual inspection and how the addition of queries impacts the clustering of the logs.

#### *6.1.3. Accuracy*

An important metric for the prediction model is its accuracy in correctly predicting labels. To assess the model's performance, we first evaluate the model's accuracy on the test set, which is a measure of how well the model generalizes to unseen data.

We use an adapted version of the *F1 score* for multi-class classification, which is the harmonic mean of multi-class precision and recall, computed with macro-averaging.[9] The F1 score with macro-averaging measures how well models perform multiclass classification over all classes, while being more punishing towards models that perform poorly on underrepresented classes.

### *6.2. Window size*

The core hyperparameter of our model is the **window size**, which determines how many log lines are fed together to the model at once. Larger window sizes allow the model to capture more context, linking log lines that are farther apart. Additionally, each log line is featured in more batches, increasing the accuracy of the majority voting system.

On the other hand, increasing the window size naturally increases the model's training time, inference time, and resource usage. We evaluated this trade-off by training the model with different window sizes and measuring the training time and resulting F1 score. The results of this experiment are shown in Table 5.

In our three-node setup, increasing $W$ beyond 50 or 60 yields no improvements in the F1 score while significantly increasing the model's training time. It must be underlined that this is a result tied to the overall size and traffic of our cluster and the dataset we used. As clusters grow larger, a larger window size may be needed to achieve the same performance. Thus, users should experiment with different window sizes to find the best trade-off between training time and accuracy for their specific use case.

[9] https://scikit-learn.org/stable/modules/generated/sklearn.metrics.f1_score.html.

**Table 6**
Inference time of the model with different window sizes.

| $W$ | Line inference time (ms) |
|---|---|
| 5 | 0.2 ms |
| 10 | 0.3 ms |
| 20 | 0.4 ms |
| 30 | 0.5 ms |
| 40 | 0.7 ms |
| 50 | 0.8 ms |
| 60 | 1.0 ms |

The model's training time scales linearly with the window size, but the size on disk does not. This is because the model's learnable parameters are not tied to the window size, and the model's size on disk remains constant, regardless of the window size: ≈ 718,000 parameters or 2.7 MB. Such a dimension is acceptable for a deep learning model that needs to perform real-time inference on data on a standard cluster; however, the performance may suffer in more resource-constrained scenarios.

The model's inference time also scales linearly with the window size, as the model needs to process more log lines at once. We define the inference time as the total time needed to process a batch of log lines, including all the data encoding and decoding, as well as the majority voting process. This time is then averaged over all the log lines in the batch, and includes the one-time overhead of loading the model in memory. We measured the inference time at the same $W$ values as the training time and using the same test dataset used for training, fed all at once. The results are shown in Table 6. Standard deviation is not reported as it is negligible in this case.

From the results, we can see that the inference time increases linearly with the window size, as we expected. The model is able to process batches of log lines very quickly: even with $W = 60$, the model processes thousands of log lines in just a few seconds. Such a performance enables K8NTEXT to be used in real-time scenarios, where logs are continuously generated and need to be processed on the fly. Given that audit logs can be already sent to a webhook, K8NTEXT can be deployed as after a webhook receiver and process logs in real time, exposing contextualized logs to users.

Finally, to ensure that the batch size does not affect these results, we repeated the experiment with different batch sizes, ranging from 500 to 5000 log lines. The results showed that for reasonable batch sizes (i.e., 2000 or more log lines, per the default K8s webhook parameters), the overhead of loading the model in memory is amortized over the entire batch, leading to a negligible difference in inference time per line (less than 0.05 ms with $B = 2000$ compared to the previous result).

To conclude, K8NTEXT shows good performance both in terms of training time and inference time, even with moderate window sizes. As re-training is needed when $W$ is changed – a limitation of the current approach – users should carefully evaluate the best window size for their use case before deploying K8NTEXT in production.

### 6.3. *Features*

The model's performance – in terms of both time and accuracy – is heavily affected by the set of chosen features. Minimizing them reduces the model size, training time, and avoids overfitting. Doing so with deep learning models, which work like *black boxes*, is unfortunately not trivial. In these cases, assessing how the model combines features is hard and approaches such as Principal Component Analysis (PCA) are unfortunately inapplicable.

We conducted a series of experiments to iteratively evaluate the importance of each feature. First, we trained the model with all features and evaluated its performance over 40 trials. Then, for each of the feature of the initial set we selected, we removed it and ran the experiment again. The results of this experiment are summarized in Fig. 6. The experiment confirmed our assumption that some features are crucial for the model's performance. In particular, some fields such as the `namespace`, `subresource` and part of the `involvedObject` field in the response, when removed, caused the most significant drop in performance. On the other hand, other fields such as the `volumeBindingMode` and `ownerReferences.controller` were found to be slightly counterproductive, as their removal led to a very slight increase in the F1 score.

### 6.4. *Train-test-validation split*

The train-test-validation split is an important driving factor in the model's performance. A good split ensures that the model generalizes well to unseen data, while a bad split can lead to overfitting and biased predictions. We evaluated the model's performance with different splits of the dataset, varying the training, testing, and validation set sizes. Fig. 7 shows the results of this experiment.

First, we set the training/testing split to 10 % and varied the training/validation split. As the latter increased, the model's performance declined, suggesting a heavy reliance of the model on a consistent amount of the dataset for its training. Indeed, the model's performance peaked at the initial 0.1 % / 0.1 % split, with a F1 score of 0.9839. This setting was chosen as the default for the model's training and all subsequent experiments.

To further investigate the model's performance, we set the training/testing split to 20 % and 30 % and varied the training/validation split again. This time, we immediately observed poor performance and decided to halt the experiment at the 30 % - 30 % split, with a F1 score of 0.9332. This further confirms our previous assumption that the model requires a large amount of data to perform optimally.

Not shown in Fig. 7 is the variance in the model's training time. As both splits increased, the model's training time decreased. This is in line with our expectations, as the model's training time scales linearly with the dataset size.

### 6.5. *Class balance*

As anticipated in Section 5.6, the dataset is inherently imbalanced, with some classes being more represented than others. This imbalance can lead to biased predictions, as the model's performance will be skewed towards the most represented classes, and thus a worse accuracy and reliability during model inference.

#### 6.5.1. *Dataset splits*

The results of Section 6.4 showed that the model's performance is heavily affected by the dataset's balance. Furthermore, a very high variance led us to investigate whether *some parts of the dataset were skewing the results*.

To assess this claim, we employed a *k-fold cross-validation* approach to verify how the accuracy varies across the dataset, with increasing values of $k$. To be even more demanding in our evaluation, we used *majority accuracy* as mentioned before. Majority accuracy takes into account how many log lines were correctly predicted, i.e., $\frac{\text{correct lines}}{\text{total lines}}$, after batches are flattened and labels are assigned back to lines. This allowed us to evaluate the model's performance on different splits of the dataset and truly obtain an accurate estimate of the model's performance on unseen data.

Fig. 8 shows the results of this experiment. Each of the bars shown in the graph represents the accuracy of the model across different k-fold splits. With $k = 10$, the model achieved a good peak accuracy (0.97) when the third split was used as the test set. On the other hand, the last split consistently performed poorly, with an accuracy between 0.65 and 0.7. The end of the dataset is dominated by CRDs and other rarely used resources, which are not well represented in the training set. This suggests that the model is biased towards the majority classes and that further work is needed to improve the model’s performance on underrepresented classes.

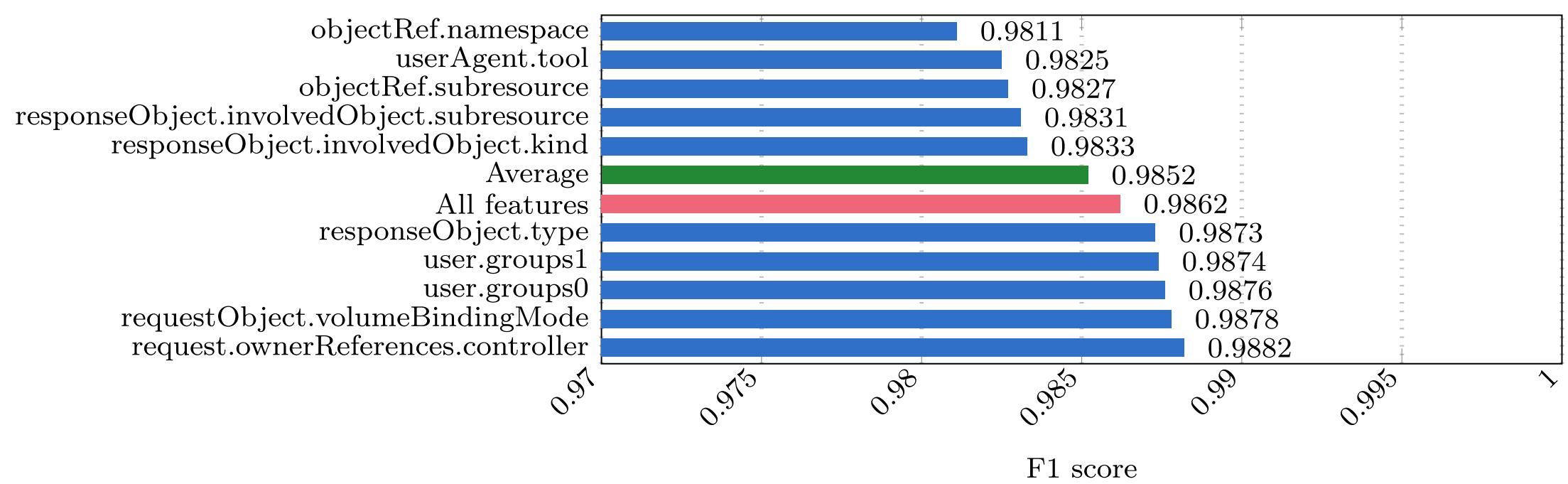

**Fig. 6.** Precision of the model with different features removed, one per time. Only the features that caused the least and most impact are shown. The central Average bar shows the average of the remaining runs that are not shown. The red “All features” bar shows the model’s performance with all features. (For interpretation of the references to colour in this figure legend, the reader is referred to the web version of this article.)

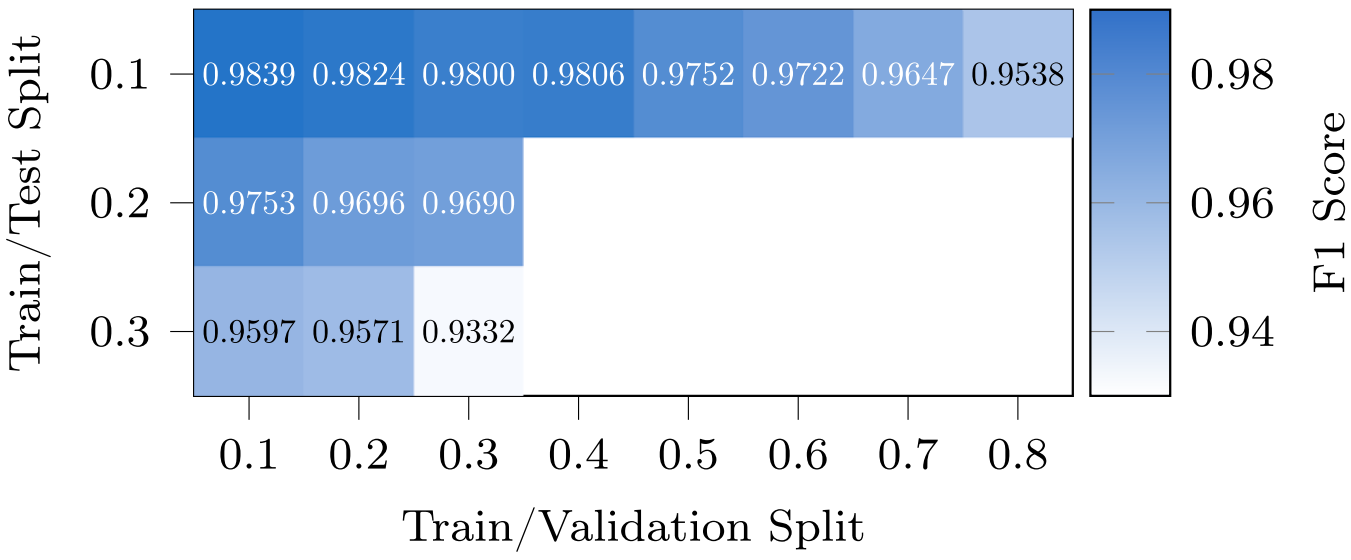

**Fig. 7.** Truncated confusion matrix of the model with different training/testing (Y-axis) and training/validation (X-axis) splits.

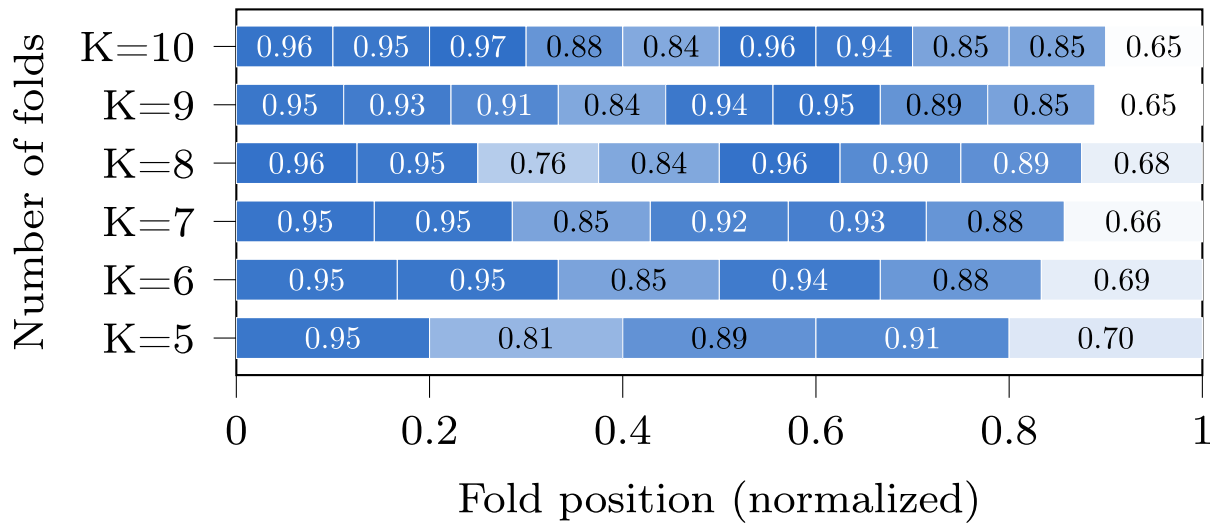

**Fig. 8.** Heatmap of the model’s accuracy with different k-fold splits. Each bar represents a set of experiments with a different $k$. Bars are each split into $k$ parts, and each bar is colored according to the accuracy of the model when that part is used as the test set. The position of splits is consistent across all bars.

### 6.5.2. *Class weights*

To further investigate the class imbalance problem, we conducted another experiment with the goal of assessing if and how the weight of each class was correlated with the model’s accuracy on that class. Fig. 9 shows a scatterplot of the model’s accuracy on each class against their weight, measured in appearances in all batches. The weight is expressed in a logarithmic scale, as some classes are extremely prevalent in the dataset, while others are very rare.

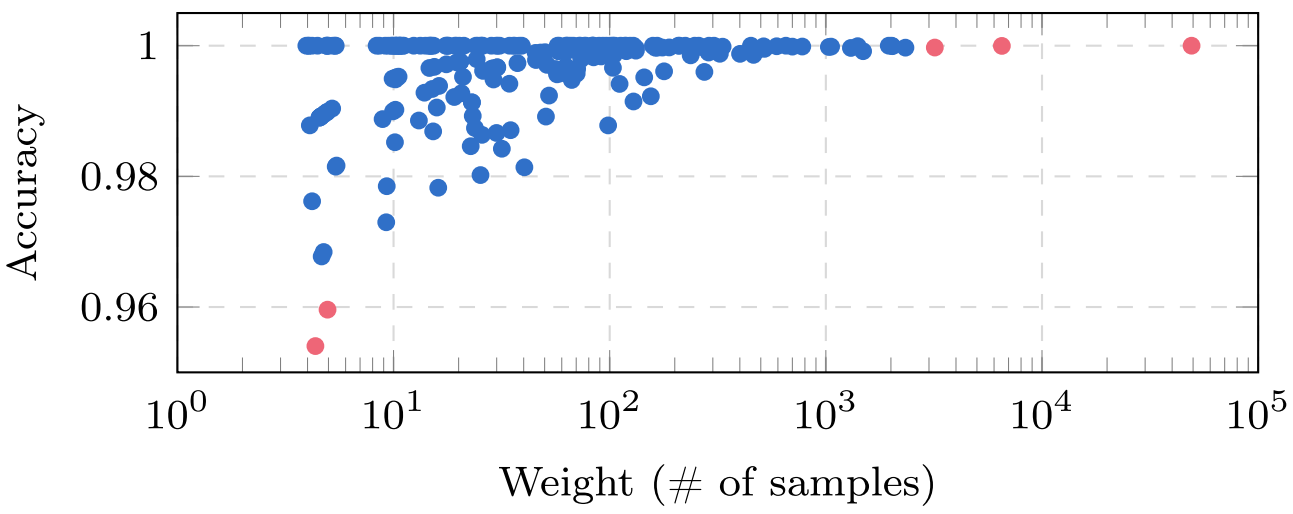

**Fig. 9.** Scatterplot of the model’s accuracy on each class against their weight (measured across all batches, logarithmic scale). The red dots are exceptional cases investigated in the text. (For interpretation of the references to colour in this figure legend, the reader is referred to the web version of this article.)

From the experiment, we discovered that the model has a two-faced behavior: on one hand, several classes consistently perform well, with an accuracy of 1.0, irrespective of their weight. The other classes’ accuracy instead scales linearly with their weight, with the most underrepresented classes having an excellent accuracy of 0.95, then nearing 1.0 as their weight in the dataset increases.

Fig. 9 also shows some exceptional cases, which are shown in red. In the top right corner, the two top red dots represent the `patch` action on Leases and the `create` action on Pods. These classes are extremely verbose, with the `patch` action on Leases accounting for $\approx 5 \cdot 10^5$ appearances in the dataset, and the `create` action on Pods being one of the most common user actions. The model performs exceptionally well on these classes, achieving an accuracy of 1.0. Indeed, most control plane traffic, being highly represented in the dataset, is always predicted with high accuracy, as the model has enough data to learn from.

On the other hand, the bottom left corner of the plot shows the worst performing classes, which are rare or hard-to-detect behaviors. These include listing LimitRanges and updating single ReplicaSets. Compared to other classes, these actions seldom appear in the dataset and have a worse accuracy. However, the model still achieves a very good accuracy of 0.95 and 0.96 respectively, which is acceptable given the low number of appearances in the dataset.

## 7. Capability and usability evaluation

This section evaluates K8NTEXT’s performance in terms of clustering effectiveness, querying time, and storage space. These metrics are important for assessing the model’s usability and efficiency in real-world scenarios. All the experiments were run under the same conditions as the previous ones, using the same virtual machine and dataset.

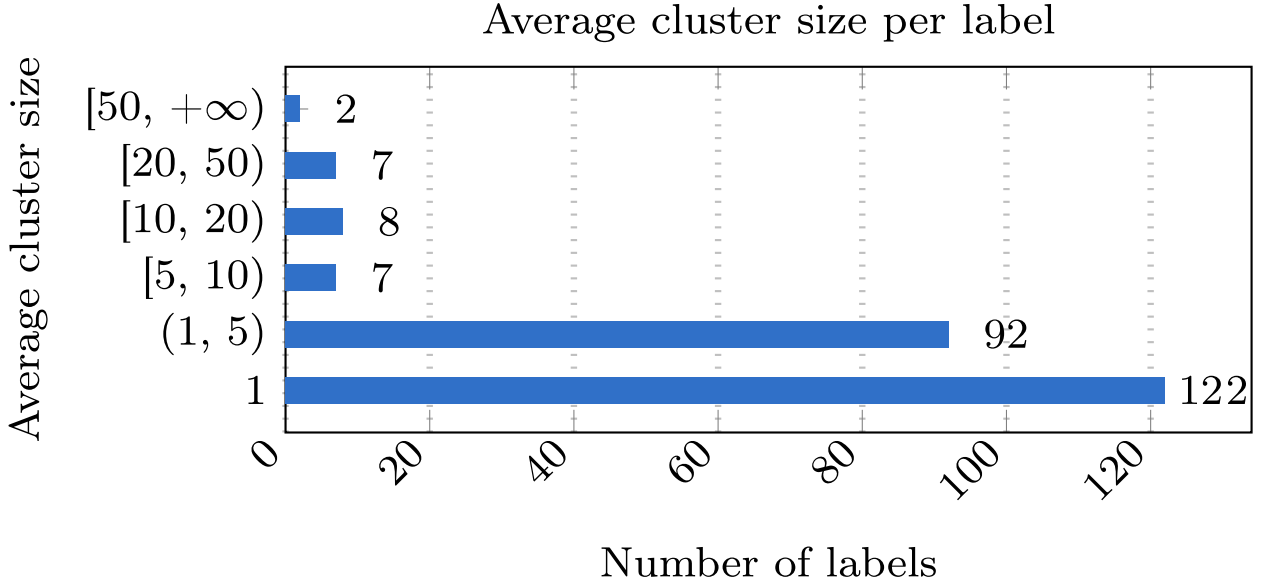

**Fig. 10.** Binned distribution of the number of log lines each label clusters together. On the Y-axis, labels are sorted into discrete intervals depending on how many log lines they cluster together. The X-axis shows how many labels fall into each bin.

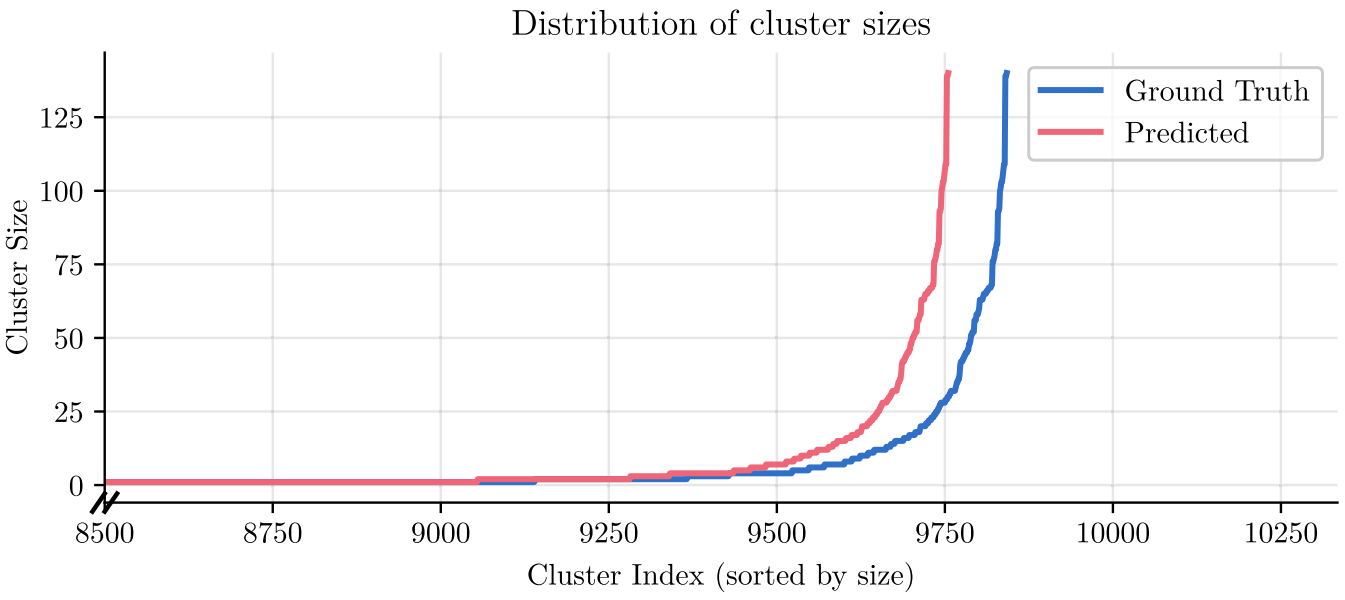

**Fig. 11.** Clustering distribution comparison of the ground truth (blue, light) and the predicted labels (red, dark). Clusters are sorted by their size, and only the last clusters are shown for brevity. The x-axis is shown starting from $x = 8500$ for readability. (For interpretation of the references to colour in this figure legend, the reader is referred to the web version of this article.)

### 7.1. Clustering

The end goal of K8NTEXT is to cut through the noise of the logs, providing users with a manageable set of contextualized events that can be used to understand the system's behavior. To achieve this, K8NTEXT clusters log lines together based on the labels assigned by the model and then further subdivides them as described in Section 5.9.

To establish a ground truth, we ran our clustering algorithm on the original labels in the dataset, which are, to the best of our knowledge, correct. The clustering algorithm yielded 9843 clusters, with an average of 1.87 log lines per cluster. Since each label produces distinct clusters, we manually inspected a subset of significant labels to ensure that the clustering algorithm was working as intended. Once we were satisfied with the results, we then evaluated the clustering performance of the algorithm both on the original labels and those predicted by the model.

The **clustering rate** is a measure of how efficiently the model can join lines together, by reducing the overall number of events that need to be inspected. Fig. 10 summarizes the results of this experiment. 92 out of 238 labels cluster 2 to 4 log lines together, representing quick and simple actions, such as listing a resource. On the other side of the spectrum, a handful of labels cluster 20 or more lines together: for example, the `update` of a Deployment. This action is a very complex one, which involves updating the underlying ReplicaSets and their Pods. Another mammoth-sized action is the deletion of a Namespace: this action is very complex and consists in deleting all the resources contained in the Namespace, which is usually hundreds of log lines. Finally, 122 labels cluster on average only a single log line, representing simple actions that do not require any context to be understood, such as the creation of a RoleBinding, or a `get` action on a Node.

By itself, the clustering rate is not a sufficient metric to evaluate K8NTEXT's performance. Thus, we also evaluated the quality of the clusters produced by the model. We ran several instances of the model using a k-fold-like approach as discussed in Section 6.5, merged the predictions, and ran the clustering algorithm on top of them. The clustering algorithm yielded 9756 clusters, 87 less than the ground truth. Of these, 9744 (99.8 %) percent were perfectly matched with the ground truth, while the remaining were mixed matches, merges, or splits. Fig. 11 shows the distribution of the two clustering results, sorted by their size.

### 7.2. Querying time

The querying time is an important metric that determines K8NTEXT's usability. We evaluated the querying time of K8NTEXT without any query and then three increasingly complex queries, on all the log lines in the dataset. Query intervals have been substituted with $t_0$ and $t_1$ for brevity. The results of this experiment are shown in Table 7.

Our experiment suggests the existence of a minimum overhead, incurred when the querying tool is used, and a linearly-scaling overhead, which increases with the complexity of the query. Given our dataset's size, we believe the querying time is acceptable for real-world scenarios, especially when programmatically polled by an external system.

### 7.3. Storage space

The impact on storage is a critical metric that determines K8NTEXT's usefulness in large-scale deployments. While the model's size on disk is relatively small, the logs can be quite large, especially in large-scale deployments. K8NTEXT gives users several options to store the logs. The raw, unprocessed logs can be stored as-is, with linking information embedded in them. Control plane information can be optionally stripped from the logs, reducing their size. Finally, the logs can be parsed and stored in a more compact format, reducing their size even further. Our dataset, amounting to 18,478 lines, originally occupied a total of 64 MB on disk. Choosing to strip down control plane information and retain the essential log details shrunk the dataset to 186 KB, resulting in three orders of magnitude reduction. This is a significant improvement in storage efficiency, but must be carefully considered in the context of data retrieval and analysis needs.

### 7.4. Applicability to other scenarios

Finally, we evaluated how well K8NTEXT performs in some alternative scenarios that differ from our training dataset. Using a deterministic scenario simulator, we generated two scenarios that mimic real-world use cases. The scenarios were designed to test K8NTEXT's ability to handle different types of workloads and user behaviors.

Scenario 1 (shown in Fig. 12) involves three users concurrently creating, deleting, and scaling Deployments. This scenario has been specif-ically chosen as a worst-case scenario for K8NTEXT because the actions performed by the users are very similar and occur in close proximity, making it difficult to distinguish between them. On the other hand, Scenario 2 (shown in Fig. 13) involves the application of NetworkPolicies and the creation of Pods to verify network connectivity.

**Table 7**
Queries used in the querying time experiment.

| ID | Query | Time (ms) |
|---|---|---|
| 0 | None | 0.073 ± 0.008 |
| 1 | `username == ` $u$ ` or stagetimestamp >= ` $t_0$ | 51.072 ± 1.364 |
| 2 | `(username == ` $u$ ` and exists(namespace)) or ` $t_0$ ` <= stagetimestamp <= ` $t_1$ | 92.800 ± 4.480 |
| 3 | `(username == ` $u$ ` and exists(namespace)) or (username == regexp(".*controller.*") and ` $t_0$ ` <= stagetimestamp <= ` $t_1$`)` | 134.558 ± 4.309 |

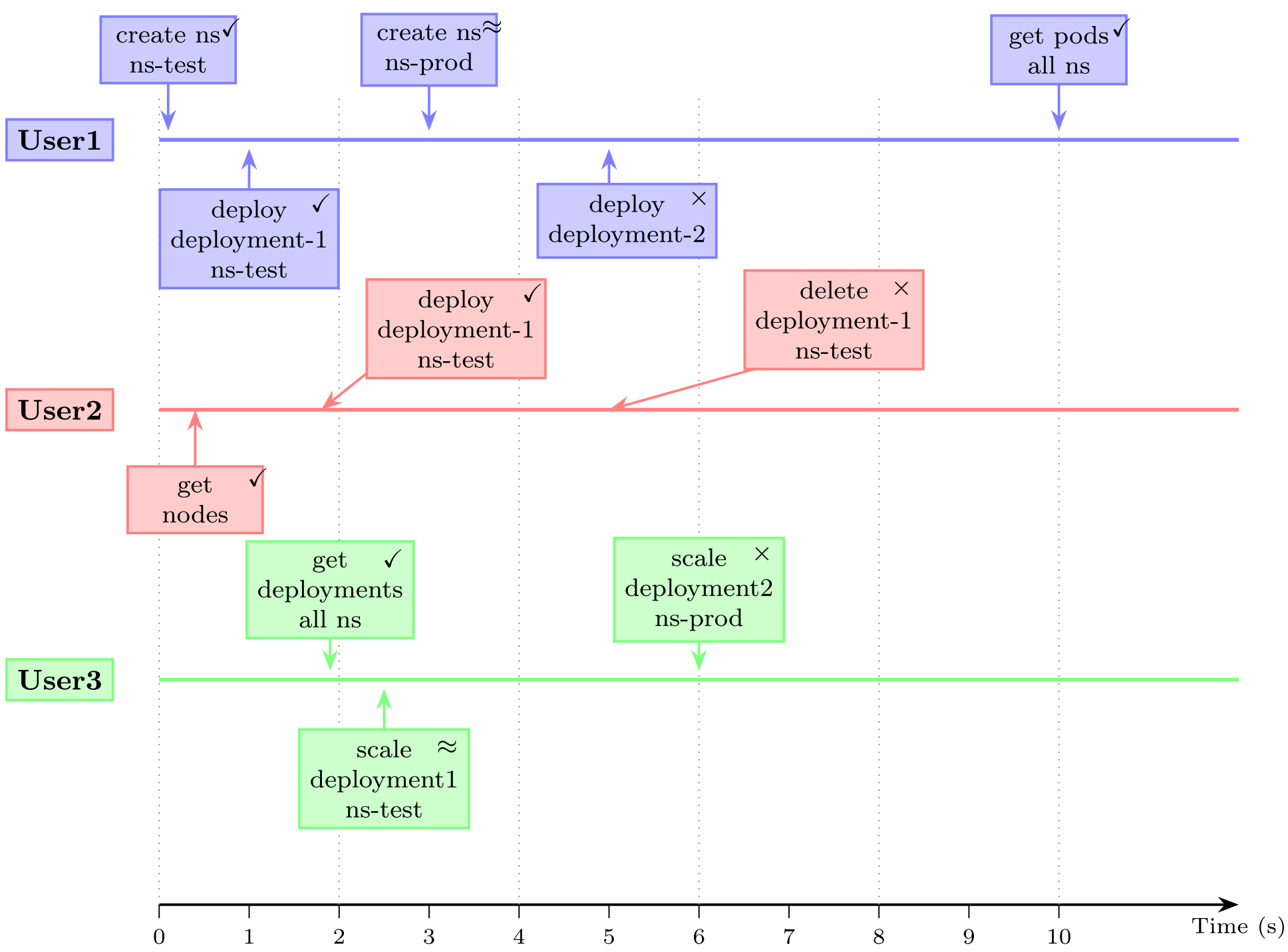

**Fig. 12.** Timeline of actions performed by different users in Scenario 1. Each color represents a different user, and each block represents an action performed by that user. `ns` is short for Namespace.

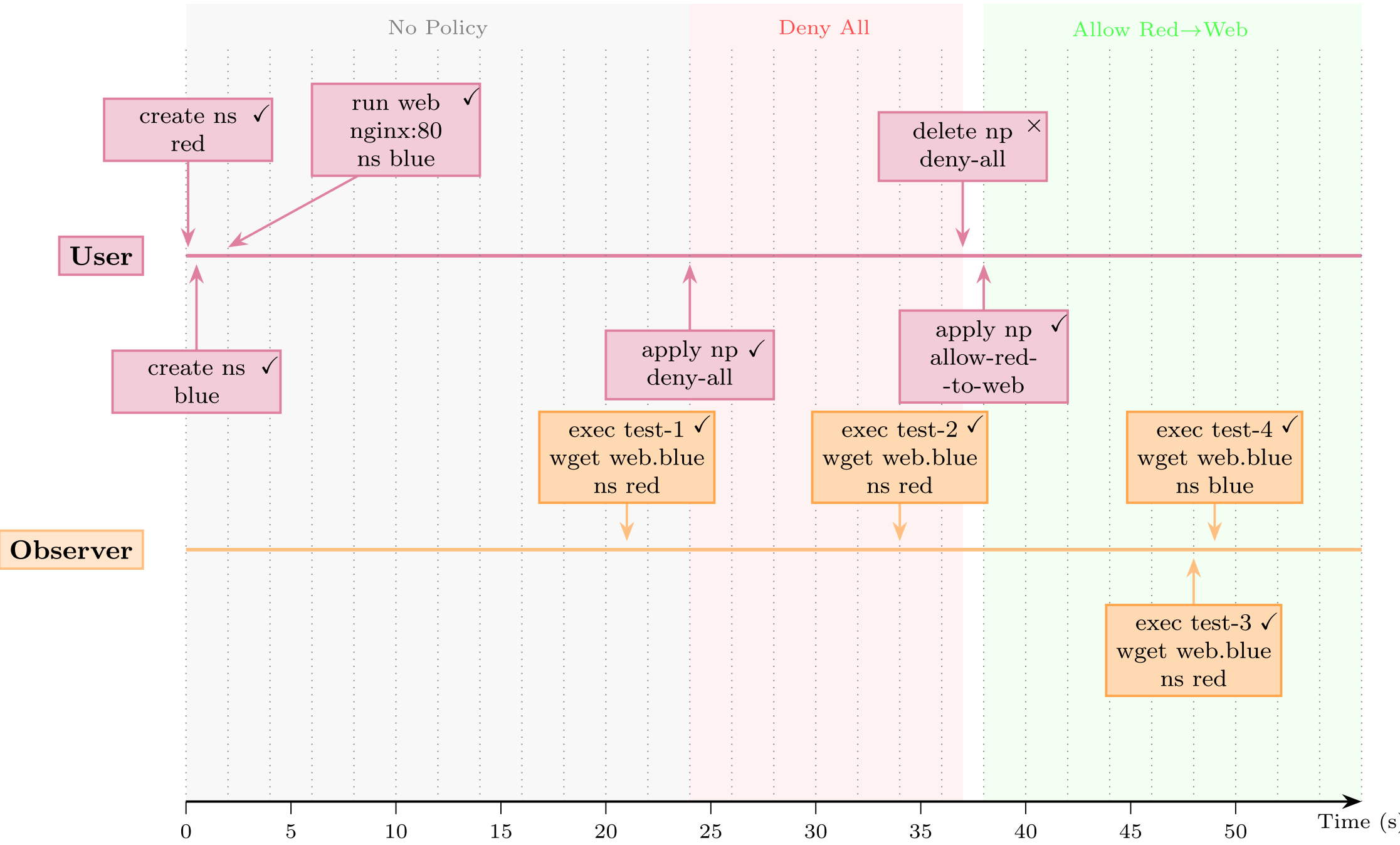

**Fig. 13.** Timeline of actions performed by different users in Scenario 2. Each color represents a different user, and each block represents an action performed by that user. `ns` is short for Namespace and `np` is short for NetworkPolicy.

Since these datasets are unlabeled, we evaluated K8NTEXT's performance by running the model, clustering, and extracting the key actions performed by each user and comparing them with the list of user actions shown in Figs. 12 and 13. In Scenario 1, of the 11 distinct actions that are performed, 6 are correctly detected (all the Deployment creations and deletions) and the context is reconstructed, 2 are correctly clusterized but not tagged as triggering actions, and 3 are mislabeled and end up in different clusters. In Scenario 2, all the actions are correctly detected and clusterized, with only one mislabeled action. Figs. 12 and 13 show a tick (✓) next to the correctly identified actions, a cross (×) next to the mislabeled actions, and an approximate sign (≈) next to lines clustered without being tagged as triggering actions.

Overall, K8NTEXT shows a good performance in these scenarios, correctly identifying and clustering most of the actions performed by the users. While there is room for improvement in terms of labeling accuracy, the clustering algorithm is able to reconstruct the context of the actions performed by the users, providing a clear and concise view of the system's behavior.

## 8. Conclusion

We presented an innovative approach for the analysis of K8s audit logs, called K8NTEXT. K8NTEXT parses, aggregates, and contextualizes raw audit logs using a ML model and a clustering algorithm, yielding a compact representation of the activities within a K8s cluster. K8NTEXT is efficient and scalable, making it suitable for real-time inspection of K8s audit logs, both by humans and machines.

As future work, we plan to explore techniques for automatic generation and labeling of K8s scenarios, as well as methods for incremental learning to adapt the model to new scenarios with minimal human intervention. Additionally, we aim to investigate the integration of K8NTEXT with other security tools to enhance monitoring and detection capabilities within K8s environments.

## CRediT authorship contribution statement

**Matteo Franzil:** Writing – original draft, Validation, Software, Project administration, Data curation; **Valentino Armani:** Visualization, Software, Resources, Conceptualization; **Luis Augusto Dias Knob:** Writing – original draft, Validation, Supervision, Methodology, Funding acquisition; **Domenico Siracusa:** Writing – original draft, Supervision, Investigation, Funding acquisition.

## Data availability

Data is available on GitHub as a citation within the paper or at https://github.com/daisyfbk/k8ntext

## Declaration of competing interest

The authors declare that they have no known competing financial interests or personal relationships that could have appeared to influence the work reported in this paper.

## Acknowledgment

This work was partially supported by project SERICS (PE00000014), MUR National Recovery and Resilience Plan funded by the European Union - NextGenerationEU.